\documentclass[a4paper,11pt]{article}
\pdfoutput=1
\usepackage[top=2.5cm, bottom=2.5cm, outer=1.5 cm, inner=1.5cm]{geometry}
\usepackage{graphicx}
\usepackage{epsfig}
\usepackage{amsfonts,amssymb,amsmath,mathrsfs}
\usepackage{cite}
\usepackage{dsfont}
\usepackage{float}
\usepackage[font=footnotesize,labelfont=bf,justification=centerlast,width=.94\textwidth]{caption}
\usepackage{authblk}
\usepackage{comment}
\usepackage{hyperref}
\usepackage[utf8]{inputenc}
\usepackage[toc,page]{appendix}
\usepackage{color,xcolor}
\usepackage{cancel}
\usepackage{tikz}
\usepackage{comment}
\usetikzlibrary{automata,positioning,arrows}
\usetikzlibrary{shapes,shadows,arrows}

\numberwithin{equation}{section}
\begin{document}



\newcommand{\red}[1]{\textcolor{red}{#1}}
\newcommand{\blue}[1]{\textcolor{blue}{#1}}
\newcommand{\gray}[1]{\textcolor{gray}{#1}} 
\newcommand{\yiannis}[1]{\textcolor{blue!70}{[#1]}} 

\newcommand{\wei}[2]{\textcolor{blue}{#1 }\todo[color=green]{WS: #2}}
\newcommand{\pankaj}[2]{\textcolor{blue}{#1 }\todo[color=yellow]{PC: #2}}
\newcommand{\boyang}[2]{\textcolor{blue}{#1 }\todo[color=orange]{BY: #2}}
\newcommand{\ioannis}[2]{\textcolor{blue}{#1 }\todo[color=cyan]{IP: #2}}

\newcommand{\plan}[1]{\begin{flushleft}
\gray{\tt *** #1 ***}
\end{flushleft}}

\def\bal#1\eal{\begin{align}#1\end{align}}

\def \bib{\bibitem}
\def\){\right)}
\def\({\left( }
\def\]{\right] }
\def\[{\left[ }

\def\nn{\nonumber}
\def\NO{\nonumber}
\def\nonu{\nonumber \\}
\def\ni{\noindent}

\def\half{\frac{1}{2}}


\newtheorem{definition}{Definition}[section]
\newtheorem{theorem}{Theorem}[section]
\newtheorem{lemma}{Lemma}[section]
\newtheorem{corollary}{Corollary}[section]
\newtheorem{proposition}{Proposition}[section]
\newtheorem{conjecture}{Conjecture}[section]


\def\a{\alpha}
\def\b{\beta}
\def\c{\chi}
\def\d{\delta}
\def\e{\epsilon}
\def\f{\phi}
\def\g{\gamma}
\def\h{\eta}
\def\j{\psi}
\def\k{\kappa}
\def\l{\lambda}
\def\m{\mu}
\def\n{\nu}
\def\om{\omega}
\def\p{\pi}
\def\th{\theta}
\def\r{\rho}
\def\s{\sigma}
\def\t{\tau}
\def\x{\xi}
\def\z{\zeta}
\def\D{\Delta}
\def\F{\Phi}
\def\G{\Gamma}
\def\J{\Psi}
\def\L{\Lambda}
\def\Om{\Omega}
\def\P{\Pi}
\def\Th{\Theta}
\def\S{\Sigma}
\def\U{\Upsilon}
\def\X{\Xi}


\def\ve{\varepsilon}
\def\vr{\varrho}
\def\vs{\varsigma}
\def\vth{\vartheta}
\def\tvf{\tilde{\varphi}}
\def\vf{\varphi}


\def\bba{\bbalpha}
\def\bbk{\bbkappa}
\def\bbg{\bbgamma}
\def\bbd{\bbdelta}
\def\bbs{\bbsigma}


\def\ca{{\cal A}}
\def\cb{{\cal B}}
\def\cc{{\cal C}}
\def\cd{{\cal D}}
\def\ce{{\cal E}}
\def\cf{{\cal F}}
\def\cg{{\cal G}}
\def\ch{{\cal H}}
\def\ci{{\cal I}}
\def\cj{{\cal J}}
\def\ck{{\cal K}}
\def\cl{{\cal L}}
\def\cm{{\cal M}}
\def\cn{{\cal N}}
\def\co{{\cal O}}
\def\cp{{\cal P}}
\def\cq{{\cal Q}}
\def\car{{\cal R}}
\def\cs{{\cal S}}
\def\ct{{\cal T}}
\def\cu{{\cal U}}
\def\cv{{\cal V}}
\def\cw{{\cal W}}
\def\cx{{\cal X}}
\def\cy{{\cal Y}}
\def\cz{{\cal Z}}


\def\bta{\textit{\textbf{a}}}
\def\btb{\textit{\textbf{b}}}
\def\btc{\textit{\textbf{c}}}
\def\btd{\textit{\textbf{d}}}
\def\bte{\textit{\textbf{e}}}
\def\btf{\textit{\textbf{f}}}
\def\btg{\textit{\textbf{g}}}
\def\bth{\textit{\textbf{h}}}
\def\bti{\textit{\textbf{i}}}
\def\btj{\textit{\textbf{j}}}
\def\btk{\textit{\textbf{k}}}
\def\btl{\textit{\textbf{l}}}
\def\btm{\textit{\textbf{m}}}
\def\btn{\textit{\textbf{n}}}
\def\bto{\textit{\textbf{o}}}
\def\btp{\textit{\textbf{p}}}
\def\btq{\textit{\textbf{q}}}
\def\btr{\textit{\textbf{r}}}
\def\bts{\textit{\textbf{s}}}
\def\btt{\textit{\textbf{t}}}
\def\btu{\textit{\textbf{u}}}
\def\btv{\textit{\textbf{v}}}
\def\btw{\textit{\textbf{w}}}
\def\btx{\textit{\textbf{x}}}
\def\bty{\textit{\textbf{y}}}
\def\btz{\textit{\textbf{z}}}

\newcommand{\sgn}{\operatorname{sgn}}

\renewcommand{\bar}{\overline}
\renewcommand{\tilde}{\widetilde}
\renewcommand{\hat}{\widehat}
\renewcommand{\leq}{\leqslant}
\renewcommand{\geq}{\geqslant}
\newcommand{\la}{\left\langle}
\newcommand{\ra}{\right\rangle}
\newcommand{\xp}{x^{+}}
\newcommand{\xm}{x^{-}}

\newcommand{\CC}{\mathbb{C}}
\newcommand{\RR}{\mathbb{R}}
\newcommand{\HH}{\mathbb{H}}
\newcommand{\ZZ}{\mathbb{Z}}
\newcommand{\cA}{\mathcal{A}}
\newcommand{\cB}{\mathcal{B}}
\newcommand{\cC}{\mathcal{C}}
\newcommand{\cD}{\mathcal{D}}
\newcommand{\cE}{\mathcal{E}}
\newcommand{\cF}{\mathcal{F}}
\newcommand{\cG}{\mathcal{G}}
\newcommand{\cH}{\mathcal{H}}
\newcommand{\cI}{\mathcal{I}}
\newcommand{\cJ}{\mathcal{J}}
\newcommand{\cK}{\mathcal{K}}
\newcommand{\cL}{\mathcal{L}}
\newcommand{\cM}{\mathcal{M}}
\newcommand{\cN}{\mathcal{N}}
\newcommand{\cO}{\mathcal{O}}
\newcommand{\cP}{\mathcal{P}}
\newcommand{\cQ}{\mathcal{Q}}
\newcommand{\cR}{\mathcal{R}}
\newcommand{\cS}{\mathcal{S}}
\newcommand{\cT}{\mathcal{T}}
\newcommand{\cU}{\mathcal{U}}
\newcommand{\cV}{\mathcal{V}}
\newcommand{\cW}{\mathcal{W}}
\newcommand{\cX}{\mathcal{X}}
\newcommand{\cY}{\mathcal{Y}}
\newcommand{\cZ}{\mathcal{Z}}


\newcommand{\bra}{\bar{a}}
\newcommand{\brb}{\bar{b}}
\newcommand{\brc}{\bar{c}}
\newcommand{\brd}{\bar{d}}
\newcommand{\bre}{\bar{e}}
\newcommand{\brf}{\bar{f}}
\newcommand{\brg}{\bar{g}}
\newcommand{\brh}{\bar{h}}
\newcommand{\bri}{\bar{i}}
\newcommand{\brj}{\bar{j}}
\newcommand{\brk}{\bar{k}}
\newcommand{\brl}{\bar{l}}
\newcommand{\brm}{\bar{m}}
\newcommand{\brn}{\bar{n}}
\newcommand{\bro}{\bar{o}}
\newcommand{\brp}{\bar{p}}
\newcommand{\brq}{\bar{q}}
\newcommand{\brr}{\bar{r}}
\newcommand{\brs}{\bar{s}}
\newcommand{\brt}{\bar{t}}
\newcommand{\bru}{\bar{u}}
\newcommand{\brv}{\bar{v}}
\newcommand{\brw}{\bar{w}}
\newcommand{\brx}{\bar{x}}
\newcommand{\bry}{\bar{y}}
\newcommand{\brz}{\bar{z}}

\newcommand{\be}{\begin{equation}}
\newcommand{\ee}{\end{equation}}
\newcommand{\bea}{\begin{eqnarray}}
\newcommand{\eea}{\end{eqnarray}}
\newcommand{\bb}{\mathbb}
\newcommand{\ba}{\begin{align}}
\newcommand{\ea}{\end{align}}
\newcommand{\bad}{\begin{aligned}}
\newcommand{\ead}{\end{aligned}}
\newcommand{\nd}{\noindent}
\newcommand{\bsub}{\begin{subequations}}
\newcommand{\esub}{\end{subequations}}
\newcommand{\beqx}{\begin{displaymath}}
\newcommand{\eeqx}{\end{displaymath}}
\newcommand{\bmat}{\left(\begin{array}}
\newcommand{\emat}{\end{array}\right)}
\newcommand*\Laplace{\mathop{}\!\mathbin\bigtriangleup}
\newcommand*\DAlambert{\mathop{}\!\mathbin\Box}



\def\Sc#1{{\hbox{\sc #1}}}      
\def\Sf#1{{\hbox{\sf #1}}}      
\def\mb#1{\mbox{\boldmath $#1$}}
\def\mf#1{\ensuremath{\mathfrak{#1}}} 
\def\bb#1{\ensuremath{\mathbb{#1}}} 


\def\slpa{\slash{\pa}}                         
\def\slin{\SLLash{\in}}                                 
\def\bo{{\raise-.3ex\hbox{\large$\Box$}}}               
\def\cbo{\Sc [}                                         
\def\pa{\partial}                                       
\def\de{\nabla}                                         
\def\dell{\nabla}                                       
\def\su{\sum}                                           
\def\pr{\prod}                                          
\def\iff{\leftrightarrow}                               
\def\conj{{\hbox{\large *}}}                            
\def\ltap{\raisebox{-.4ex}{\rlap{$\sim$}} \raisebox{.4ex}{$<$}}   
\def\gtap{\raisebox{-.4ex}{\rlap{$\sim$}} \raisebox{.4ex}{$>$}}   
\def\face{{\raise.2ex\hbox{$\displaystyle \bigodot$}\mskip-2.2mu \llap {$\ddot
        \smile$}}}                                   
\def\dg{\dagger}                                     
\def\ddg{\ddagger}                                   
\def\trans{\mbox{\scri T}}                           
\def\>{\rangle}                                      
\def\<{\langle}                                      


\def\tx#1{\text{#1}}
\def\sp#1{{}^{#1}}                                   
\def\sb#1{{}_{#1}}                                   
\def\sptx#1{{}^{\rm #1}}                           
\def\sbtx#1{{}_{\rm #1}}                           
\newcommand{\sub}[1]{\phantom{}_{(#1)}\phantom{}}    
\def\oldsl#1{\rlap/#1}                               
\def\slash#1{\rlap{\hbox{$\mskip 1 mu /$}}#1}        
\def\Slash#1{\rlap{\hbox{$\mskip 3 mu /$}}#1}        
\def\SLash#1{\rlap{\hbox{$\mskip 4.5 mu /$}}#1}      
\def\SLLash#1{\rlap{\hbox{$\mskip 6 mu /$}}#1}       
\def\wt#1{\widetilde{#1}}                            
\def\Hat#1{\widehat{#1}}                             
\def\lbar#1{\ensuremath{\overline{#1}}}              
\def\bra#1{\left\langle #1\right|}                   
\def\ket#1{\left| #1\right\rangle}                   
\def\VEV#1{\left\langle #1\right\rangle}             
\def\abs#1{\left| #1\right|}                         
\def\leftrightarrowfill{$\mathsurround=0pt \mathord\leftarrow \mkern-6mu
        \cleaders\hbox{$\mkern-2mu \mathord- \mkern-2mu$}\hfill
        \mkern-6mu \mathord\rightarrow$}        
\def\dvec#1{\vbox{\ialign{##\crcr
        \leftrightarrowfill\crcr\noalign{\kern-1pt\nointerlineskip}
        $\hfil\displaystyle{#1}\hfil$\crcr}}}           
\def\dt#1{{\buildrel {\hbox{\LARGE .}} \over {#1}}}     
\def\dtt#1{{\buildrel \bullet \over {#1}}}              
\def\der#1{{\pa \over \pa {#1}}}                        
\def\fder#1{{\d \over \d {#1}}}                         
\def\tr{{\rm tr \,}}                                    
\def\Tr{{\rm Tr \,}}                                    
\def\diag{{\rm diag \,}}                                
\def\Re{{\rm Re\,}}                                     
\def\Im{{\rm Im\,}}                                     
\def\mrp{\mathrm{p}}

\def\partder#1#2{{\partial #1\over\partial #2}}        
\def\parvar#1#2{{\d #1\over \d #2}}                    
\def\secder#1#2#3{{\partial^2 #1\over\partial #2 \partial #3}}  
\def\on#1#2{\mathop{\null#2}\limits^{#1}}              
\def\bvec#1{\on\leftarrow{#1}}                         
\def\oover#1{\on\circ{#1}}                             


\def\Deq#1{\mbox{$D$=#1}}                               
\def\Neq#1{\mbox{$cn$=#1}}                              
\newcommand{\ampl}[2]{{\cal M}\left( #1 \to #2 \right)} 


\def\NPB#1#2#3{Nucl. Phys. B {\bf #1} (19#2) #3}
\def\PLB#1#2#3{Phys. Lett. B {\bf #1} (19#2) #3}
\def\PLBold#1#2#3{Phys. Lett. {\bf #1}B (19#2) #3}
\def\PRD#1#2#3{Phys. Rev. D {\bf #1} (19#2) #3}
\def\PRL#1#2#3{Phys. Rev. Lett. {\bf #1} (19#2) #3}
\def\PRT#1#2#3{Phys. Rep. {\bf #1} C (19#2) #3}
\def\MODA#1#2#3{Mod. Phys. Lett.  {\bf #1} (19#2) #3}


\def\norder{\raisebox{-.13cm}{\ensuremath{\circ}}\hspace{-.174cm}\raisebox{.13cm}{\ensuremath{\circ}}}
\def\bz{\bar{z}}
\def\bw{\bar{w}}
\def\-{\hphantom{-}}
\newcommand{\dd}{\mbox{d}}
\newcommand{\scr}{\scriptscriptstyle}
\newcommand{\scri}{\scriptsize}
\def\rand#1{\marginpar{\tiny #1}}               
\newcommand{\rstar}{\rand{\bf\large *}}
\newcommand{\rup}{\rand{$\uparrow$}}
\newcommand{\rdown}{\rand{$\downarrow$}}


\title{\textbf{SYK Model in a Non-Gaussian disorder ensemble and emergent Coleman's mechanism}}

\author[a]{Ido Ben-Dayan\thanks{\noindent E-mail:~ido.bendayan@gmail.com }}
\author[a,b]{Pankaj Chaturvedi \thanks{\noindent E-mail:~  cpankaj1@gmail.com}}

\affil[a]{
\textit{Physics Department, Ariel University, Ariel 4070001, Israel}}

\affil[b]{
\textit{Department of Physics, NIT Silchar, Assam  788010, India}}

\maketitle
\thispagestyle{empty}

\abstract{We consider the case of the SYK model with non-gaussian disorder in the large $N$ limit. After obtaining the effective action, we derive the density of states and the free energy of the modified theory. We show that the non-gaussian disorder corresponds to a non-local Liouville theory, and non-minimally coupled 2D gravity action. It also provides a nice realization of Colemania - Coleman's idea from the 80s of generating a small Cosmological Constant. Finally, we also calculate out of time order correlation functions (OTOC) for the model.
 }

\newpage
\tableofcontents
\addtocontents{toc}{\protect\setcounter{tocdepth}{2}}
\setcounter{page}{1}

\section{Introduction}\label{sec:intro}
 The Sachdev-Ye-Kitaev (SYK) model \cite{Sachdev:1993PhRvL, Sachdev:2010uj, Parcollet:1999oa} is a remarkable theoretical construct that lies at the intersection of condensed matter physics, quantum field theory, and high-energy physics. The SYK model consists of a collection of N-Majorana fermions with polynomial all to all interactions of order-q involving coupling constants described by a Gaussian random variable. In the strong coupling regime it shows a non-Fermi liquid behavior. This means that the traditional concepts of quasiparticles and Landau's Fermi liquid theory do not apply, and the system displays unconventional transport and thermodynamic properties\cite{Davison:2016ngz}. This model and its variants \cite{Gu:2016oyy, Witten:2016iux, Fu:2016vas, Klebanov:2016xxf, Gross:2016kjj, Krishnan:2016bvg, Turiaci:2017zwd, Murugan:2017eto, Davison:2016ngz, Chen:2017dav, Jian:2017unn, Cai:2017vyk, Peng:2017kro, Maldacena:2018lmt} exhibit several interesting properties at low energies and temperatures. Specifically, it becomes solvable in the large N ($N\to \infty$) limit. It exhibits maximally chaotic behavior which is captured by its out-of-time-ordered correlation functions (OTOCs) \cite{Maldacena:2015waa, Kitaev:2015tk, Maldacena:2016hyu, Maldacena:2016upp, Polchinski:2016xgd, Bagrets:2017pwq, 2021Symm...13..599C}, an emergent reparametrization symmetry and its subsequent breaking to $SL(2, R)$ both explicitly and spontaneously, low energy dynamics being governed by $SL(2, R)$ invariant Schwarzian action and so on. Despite its chaotic nature, the SYK model tends to thermalize, aligning with the Eigenstate Thermalization Hypothesis (ETH)\cite{Sonner:2017hxc}. This suggests that the model can reach thermal equilibrium despite its intricate dynamics. The entanglement properties of the SYK model have been explored extensively \cite{Liu:2017kfa, Huang:2017nox, Zhang:2020sci, Zhang:2022yaw, Haldar:2020ymg}, revealing connections to various concepts in quantum information theory and providing insights into the nature of quantum entanglement and its role in chaotic systems. 
 
Recently the investigation of SYK models and their connection to two-dimensional gravity \cite{Kitaev:2015tk, Sachdev:2010um, Almheiri:2014cka, Sachdev:2015efa, Maldacena:2016hyu, Maldacena:2016upp, Engelsoy:2016xyb,  Jensen:2016pah, Kitaev:2017awl} has opened a new paradigm within the framework of gauge/gravity duality. This duality is motivated by the fact that the symmetry-breaking pattern, namely the emergence of the reparametrization symmetry in the infrared (IR) regime and its subsequent breaking to $SL(2, R)$ both explicitly and spontaneously, is captured both by the Jackiw-Teitelboim (JT) gravity \cite{  Kitaev:2015tk, Maldacena:2016hyu, Maldacena:2016upp} and the SYK model. In other words, the low-energy effective actions of the SYK model and the JT model are both described by a Schwarzian derivative \cite{  Kitaev:2015tk, Maldacena:2016hyu, Maldacena:2016upp} which governs the dynamics of a pseudo-Goldstone mode associated with the reparametrization invariance. Moreover, the entropy difference between near-extremal black holes and extremal ones is linear in the Hawking temperature, resembling the linear growth of entropy at low temperatures in the SYK model. 

 SYK models have also found connections with random matrix theories (RMT) and their classifications. RMT in general provide a theoretical framework for understanding statistical properties of complex systems that exhibit chaotic behavior, such as energy levels of nuclei, correlations in disordered systems, and even properties of some physical systems exhibiting quantum chaos\cite{mehta2004random, Wigner:791e55682183}. In RMT, the elements of a random matrix are often assumed to be  independent and identically distributed Gaussian random variables. Depending upon whether the random matrix has real, complex or quaternionic elements, the RMT can be classified under three main Gaussian ensembles: the Gaussian Orthogonal Ensemble (GOE), the Gaussian Unitary Ensemble (GUE), or the Gaussian Symplectic Ensemble (GSE) respectively. RMT also show a universal behavior which refers to the remarkable phenomenon where the eigenvalue distributions of matrices from different ensembles converge to the same universal distribution as the matrix size becomes very large. The connection between the SYK model and RMT lies in the spectral properties and statistical behavior of these systems \cite{You_2017, Garcia-Garcia:2016mno, Dyer:2016pou, Cotler:2016fpe}. In particular, the late-time behavior of the disorder averaged thermal spectral form factor for the SYK model tends to a plateau characterizing the discreteness of the spectrum. Remarkably, the spectral form factor for the random matrix theory corresponding to Gaussian unitary, orthogonal, and symplectic ensembles, mimics the same late-time behavior as observed for the case of the SYK model \cite{ Dyer:2016pou, Cotler:2016fpe}. One interesting extension of RMT is its tensor generalization, which involves random tensors instead of random matrices \cite{Gurau_2016, Bonzom:2012hw, Gurau:2011kk}. Such generalizations are called random tensor theories (RTT) and mimic all the properties of RMT discussed above.

Motivated by the above dualities, in this work we have considered the SYK model with interacting Majorana fermions where the values of the coupling constants are now picked from a non-Gaussian distribution. We will call the model considered here as non-Gaussian SYK throughout the manuscript. The idea behind considering such a model is that different distributions of couplings one must recover the usual SYK results in the large N limit with $\cO\left(1/N\right)$ or higher corrections that might differ for each such distribution. Moreover, for such models it is expected that different distributions of couplings will give rise to different 2D gravity theories at low energy, generating a whole family of such theories that can be analyzed at the quantum level. The specific model we consider here, is inspired by the p-spin glass model with non-Gaussian, correlated disorder studied in \cite{Bonzom_2013} and the SYK model containing flavored complex fermions with non-Gaussian disorder averaging over the random couplings studied in \cite{Krajewski:2018lom}. The construction of such models is motivated by the Gaussian universality of the random tensor model which was first established in \cite{Gurau:2011kk}. In random tensor model, the elements of a random tensor of rank D are often assumed to be independent and identically distributed complex random variables. The authors in \cite{Gurau:2011kk} showed that in the large N limit the distribution of $N^D$ elements of a random tensor of rank D converges to that of a Gaussian tensor model. They also established that if one considers a perturbed Gaussian distribution for the elements of a random tensor then in the same large N limit it again reduces to a Gaussian distribution with the covariance now being dependent on the nature of the perturbation. 

It is clear from the above discussion that introducing non-Gaussian disorder averaging has interesting consequences on the properties of the SYK model. In this work we have explored several interesting constructs like the partition function, time ordered and out of time ordered correlation functions, and the free energy of the non-Gaussian SYK model. The main observations of this study can be summarized as follows:

\begin{enumerate}
    \item In the large N limit we have obtained the disorder averaged bilocal effective action for the non-Gaussian SYK model. It is observed that this non-Gaussian disorder averaging for the SYK model with interactions of order-q results in a modification of the variance of the Gaussian distribution of couplings to the leading order in N. One also observes the emergence of an $\cO\left(1/N^{q-2}\right)$ non-local term in the large N effective action. Similar observation was also made in the case of the SYK model containing flavored complex fermions with non-Gaussian disorder averaging in \cite{Krajewski:2018lom}. 

    \item In the large N and large q limit (where q is the order of interactions) the effective action for the non-Gaussian SYK model is described by a nonlocal Liouville field theory for the fluctuations around the free fermion Green's function. The two dimensional euclidean Liouville field theory contains a cosmological term and a subdominant nonlocal term. This theory is very similar to the Eisntein gravity with positive cosmological constant and a nonlocal term considered by Veneziano \cite{Veneziano:1989hd} in a four dimensional setting. The idea of considering such a nonlocal euclidean quantum gravity theory with wormholes solutions was inspired by Coleman in \cite{Coleman:1988tj}. Using such a theory he further argued that if the wormholes exist, they have the effect of making the cosmological constant small. Polchinski showed that the method works only at space-time dimensions $D=2\,(\text{mod} \, 4)$ \cite{Polchinski:1988ua}, which is the case at hand. Hence, using the approach adopted in \cite{Veneziano:1989hd} we show that one can also have a large volume and a small cosmological contant solution to the two dimensional nonlocal euclidean Liouville field theory corresponding to the non-Gaussian SYK model, providing a nice realization of Colemania.

    \item In the low temperature regime, we have also computed various physical quantities pertaining to the non-Gaussian SYK model namely: the thermal partition function, density of states, time and out of time ordered correlation functions and the Lyapunov exponent. The results indicate that all of these quantities to the leading order in the large N expansion matches with the usual SYK model with Gaussian disorder averaging. We also show that the non-Gaussianity of the SYK model does not effect the Lyapunov exponent obtained from its four-point OTOC indicating that the model is still maximally chaotic.

\end{enumerate}

The paper is organized as follows:  In Section 2, we introduce the SYK model with non-Gaussian disorder and in the large N limit, and obtain the bilocal effective action in Euclidean time. In this section we also outline the procedure for computing the thermal partition and correlation function for the non-Gaussian SYK model which involves using a Hubbard-Stratonovich transformation. In section 3, in the large N and large q limit we derive the effective action for the non-Gaussian SYK model which is described by a nonlocal Liouville field theory. We also compute the free energy for the non-Gaussian SYK model in this regime and elucidate on the possible Coleman mechanism for nonlocal Liouville field theory. In section 4, we derive the nonlocal Schwarzian theory corresponding to the non-Gaussian SYK model in the low temperature regime. Here we also compute the four-point OTOC and the corresponding Lyapunov exponent for the non local Schwarzian theory corresponding to the non-Gaussian SYK model. We then conclude,
and discuss the implications of introducing non-Gaussian disorder in the SYK model and  possible open problems.

\section{SYK model with non-Gaussian disorder}
As discussed in the previous section the theory of interest here is the one-dimensional SYK model with N Majorana fermions described by the following Euclidean action

\bal
S(\psi)=\int d\tau\(\frac{1}{2}\sum\limits_{i=1}^{N}\psi_{i}\pa_{\t}\psi_{i}+\frac{(i)^{q/2}}{q!}\sum\limits_{\{i_a\}}^{N} J_{i_1i_2\cdots i_q}\psi_{i_1}\psi_{i_2}\cdots \psi_{i_q}\),\label{SYKAction}
\eal
where the label set $\{{i_a\}}:=\{i_1,i_2,\cdots,i_q\}$ such that $\{1\leq i_1<i_2<\cdots <i_{q}\leq N\}$, and $q$ (an even number) is the order of the interaction term involving q-fermions at a time. Here, a factor of $i$ is necessary to make the Hamiltonian hermitian when $q=2 \mod{(4)}$. It may also be seen that the SYK model is not time reversal symmetric for odd $q/2$. Thus we restrict ourselves to the case when $q$ is a multiple of four as these models have time reversal symmetric the interactions which are dominant at low energy. The rank q tensor $J_{i_1i_2\cdots i_q}$ in the SYK action plays the role of a coupling constant. Each of the components of the rank q tensor is defined conventionally as a Gaussian random variable whose values are picked from the distribution
\bal
P(J_{i_1 i_2\cdots i_q})=\sqrt{\frac{N^{q-1}}{2\pi\;\s^2}} \exp{\(-N^{q-1}\frac{J^2_{i_1i_2\cdots i_q}}{2\s^2}\)},\label{Gdis}
\eal
having zero mean and the variance
\bal
\VEV{J^2_{i_1i_2\cdots i_q}}=\frac{J^2\;(q-1)!}{N^{q-1}}=\frac{2^{q-1}\;(q-1)!}{q\;N^{q-1}}\cJ^2=\frac{\s^2}{N^{q-1}},\label{varSYK}
\eal
characterized by the dimension one parameter $J$ which is the same for all the components. The SYK model described here is subjected to a quenched disorder which implies that the free energy is obtained as an average over the couplings. The disorder averaging can be defined in the following way
\bal
\VEV{f(J,\psi)}_{J}=\frac{\int\prod\limits_{\{i_a\}} dJ_{i_1i_2\cdots i_q}\;f(J,\psi)\;\exp{\(-N^{q-1}\sum\limits_{\{i_a\}}^{N}\frac{J^2_{i_1i_2\cdots i_q}}{2\s^2}\)}}{\int\prod\limits_{\{i_a\}} dJ_{i_1i_2\cdots i_q}\;\exp{\(-N^{q-1}\sum\limits_{\{i_a\}}^{N}\frac{J^2_{i_1i_2\cdots i_q}}{2\s^2}\)}},\label{Gavg}
\eal
where $f(J,\psi)$ is an arbitrary function of the couplings and the fermionic fields.

Given the Gaussian disorder averaging in \eqref{Gavg}, one may ask whether it is possible to consider an SYK model with more general disorder averaging. For this, we consider that the random couplings $J_{i_1i_2\cdots i_q}$ take values according to a more general distribution which may be given as
\bal
P(J_{i_1 i_2\cdots i_q})&=C\exp{\[-N^{q-1}\(\sum\limits_{\{i_a\}}^{N}\frac{J^2_{i_1i_2\cdots i_q}}{2\s^2}+V(J_{i_1i_2\cdots i_q})\)\]},\nn\\
V(J_{i_1i_2\cdots i_q})&=-\frac{\l}{4 N \s^4} \(\sum\limits_{\{i_a\},\{k_a\}}^{N}J^2_{i_1i_2\cdots i_q}J^2_{k_1k_2\cdots k_q}\),\label{GDdis}
\eal
where $V(J_{i_1i_2\cdots i_q})$ denotes the deviation from a Gaussian distribution. Notice that now the probability distribution is a complicated non-gaussian multivariate function, rather than a product of independent gaussians. The division by $\s^4$ in the expression of $V(J_{i_1i_2\cdots i_q})$ is done to make the quartic term scale independent and hence $\l$ is a dimensionless parameter here. Moreover, the division of the coupling constant $\l$ by $N$ is done to make the quartic interaction strength comparable to the quadratic term in the action at large $N$. Such a division also results in a coupling constant $\lambda$ that can in principle be large and is independent of $N$. The case of an SYK model for complex fermions with flavor studied in \cite{Krajewski:2018lom} considers a coupling constant $\lambda$. The results here agree with their analysis if we take their $\lambda\rightarrow \lambda/N$ and consider SYK model with Majorana fermions. The disorder averaging in \eqref{Gavg} can now be modified as follows
\bal
\VEV{f(J,\psi)}_{J}=\frac{\int \prod\limits_{\{i_a\}} dJ_{i_1i_2\cdots i_q}\;f(J,\psi)\;\exp{\[-N^{q-1}\(\sum\limits_{\{i_a\}}^{N}\frac{J^2_{i_1i_2\cdots i_q}}{2\s^2}+\frac{\l}{N}\left(\sum\limits_{\{i_a\}}^{N}\frac{J^2_{i_1i_2\cdots i_q}}{2 \s^2}\right)^2\)\]}}{\int \prod\limits_{\{i_a\}} dJ_{i_1i_2\cdots i_q}\;\exp{\[-N^{q-1}\(\sum\limits_{\{i_a\}}^{N}\frac{J^2_{i_1i_2\cdots i_q}}{2\s^2}+\frac{\l}{N}\left(\sum\limits_{\{i_a\}}^{N}\frac{J^2_{i_1i_2\cdots i_q}}{2 \s^2}\right)^2\)\]}},\label{GDavg}
\eal
and on using the Hubbard-Stratonovich transformation it can be further written down as 
\bal
\VEV{f(J,\psi)}_{J}=\frac{\int\cD J\;dM\;f(J,\psi)\;\exp{\[-\(\frac{M^2}{2}+N^{q-1}\(1+i\(\frac{2\l }{N^{q}}\)^{1/2}M \)\sum\limits_{\{i_a\}}^{N}\frac{J^2_{i_1i_2\cdots i_q}}{2\s^2}\)\]}}{\int\cD J\;dM\;\exp{\[-\(\frac{M^2}{2}+N^{q-1}\(1+i\(\frac{2\l }{N^{q}}\)^{1/2}M \)\sum\limits_{\{i_a\}}^{N}\frac{J^2_{i_1i_2\cdots i_q}}{2\s^2}\)\]}}.\label{GDJMavg}
\eal
where $M$ is some auxiliary parameter. It is to be noted that in \eqref{GDJMavg} integration over each $M$ contributes a $\sqrt{2\pi}$ factor which cancels with the same factor coming from the denominator. The main focus of giving the formulation in \eqref{GDJMavg} for a disorder averaging is to evaluate the free energy $F(J)$ of the SYK model. This is usually achieved by employing the replica trick 
\bal
F(J)=-\frac{1}{\b}\VEV{\ln{Z(J)}}_{J}=-\frac{1}{\b}\lim_{n\to 0}\frac{\VEV{Z^{n}(J)}_{J}-1}{n},\label{reptrick}
\eal
where
\bal
Z^{n}(J)=\int \prod\limits_{i}d\psi^{\a}_{i}\;\exp{\[-\int d\tau\;\sum\limits_{\a=1}^{n}\(\frac{1}{2}\sum\limits_{i=1}^{N}\psi^{(\a)}_{i}\pa_{\t}\psi^{(\a)}_{i}+\frac{(i)^{q/2}}{q!}\sum\limits_{\{i_a\}}^{N} J_{i_1i_2\cdots i_q}\psi^{(\a)}_{i_1}\psi^{(\a)}_{i_2}\cdots \psi^{(\a)}_{i_q}\)\]},\label{Zrep}
\eal
and $\a=\{1,\cdots,n\}$ is the replica index. The relation in \eqref{reptrick} further implies that one only needs to evaluate the quantity $\VEV{Z^{n}(J)}_{J}$ and then take the replica limit to evaluate the disorder averaged free energy. Thus considering \eqref{GDJMavg}, the disorder average of the replicated partition function in \eqref{Zrep} can be written as
\bal
\VEV{Z^{n}(J)}_{J}&=\int\;\cD J\;\cD \psi\;dM\;\exp\left[-\frac{1}{2}\int d\t \sum\limits_{\a=1}^{n}\sum\limits_{i=1}^{N}\psi^{(\a)}_{i}\pa_{\t}\psi^{(\a)}_{i}-\frac{M^2}{2}\right.\nn\\
&\left. -\(N^{q-1}\(1+i\(\frac{2\l}{N^q}\)^{1/2} M\)\sum\limits_{\{i_a\}}^{N}\frac{J^2_{i_1i_2\cdots i_q}}{2\s^2}\right.\right.\nn\\
&
\left.\left.-\sum\limits_{\{i_a\}}^{N}\(\frac{(i)^{q/2}}{q!}\int d\t\;\sum\limits_{\a=1}^{n} \psi^{(\a)}_{i_1}\psi^{(\a)}_{i_2}\cdots \psi^{(\a)}_{i_q}\)J_{i_1i_2\cdots i_q}\)\right],\label{Znrep0}
\eal

Having obtained the replica partition function in \eqref{Znrep0}, one can now perform the disorder average to obtain the following
\bal
\VEV{Z^{n}(J)}_{J}=&\int \cD\psi\;dM\;\left(\sqrt{\frac{2\pi  N^{1-q}\s^2}{\(1+i\sqrt{\frac{2\l}{N^q}}\;M\)}}\right)^{\binom{N}{q}}\exp\left[n\left(-\frac{1}{2n}\int d\t \sum\limits_{\a=1}^{n}\sum\limits_{i=1}^{N}\psi^{(\a)}_{i}\pa_{\t}\psi^{(\a)}_{i}-\frac{M^2}{2}\right.\right.\nn\\
&\left.\left.+\frac{1}{2(q!)^2}\int \int d\t d\t'\sum\limits_{\{i_a\}}^{N} \frac{\(\sum\limits_{\a=1}^{n} \psi^{(\a)}_{i_1}(\t)\cdots \psi^{(\a)}_{i_q}(\t) \)\(\sum\limits_{\b=1}^{n} \psi^{(\b)}_{i_1}(\t')\cdots \psi^{(\b)}_{i_q}(\t') \)}{  N^{q-1}\(\frac{1}{\s^2}+i\sqrt{\frac{2\l}{N^q}}\;\frac{ M}{\s^2}\)}\right)\right]\nn\\
= & \(\frac{2\pi \s^2}{ N^{q-1}}\)^{\frac{1}{2}\binom{N}{q}}\int \cD\psi\;dM\;\exp\left[\left(-\frac{1}{2}\int d\t \sum\limits_{\a=1}^{n}\sum\limits_{i=1}^{N}\psi^{(\a)}_{i}\pa_{\t}\psi^{(\a)}_{i}\right.\right.\nn\\
&\left.\left.-\frac{M^2}{2}-\frac{1}{2}\binom{N}{q}\ln{\(1+i\sqrt{\frac{2\l}{N^q}}\;M\)}\right.\right.\nn\\
&\left.\left.+\frac{N\s^2}{2~q!\(1+i\sqrt{\frac{2\l}{N^q}}M\)}\int \int d\t d\t'\sum\limits_{\a,\b}^{n}\(\frac{1}{N}\sum\limits_{i=1}^{N}\psi^{(\a)}_{i}(\t)\psi^{(\b)}_{i}(\t')\)^{q}\right)\right],\label{ZnJavg}
\eal
where from the last line in the above expression it is clear that the replica partition function can not be further simplified and one has to consider certain approximations which we describe next.

\paragraph{Replica partition function in small $\l$ and Large N approximation:}

Using the formal expansion
\bal
&\(1+i\sqrt{\frac{2\l}{N^q}}\;M\)^{-1}=\(1+\sum\limits_{k=1}^{\infty}(-1)^{k}\(i\sqrt{\frac{2\l}{N^q}}\;M\)^{k}\),\label{Bexp}\\
&-\frac{1}{2}\ln\(1+i\sqrt{\frac{2\l}{N^q}} M\)=\frac{1}{2}\sum\limits_{k=1}^{\infty}\frac{(-1)^{k}}{k}\(i\sqrt{\frac{2\l}{N^q}}\;M\)^{k},\label{Aexp}
\eal
and $\binom{N}{q}\simeq N^q$, one can write \eqref{ZnJavg} as
\bal
\VEV{Z^{n}(J)}_{J}=& \int \cD\psi\;dM\;\exp\left[\left(-\frac{1}{2}\int d\t \sum\limits_{\a=1}^{n}\sum\limits_{i=1}^{N}\psi^{(\a)}_{i}\pa_{\t}\psi^{(\a)}_{i}-\frac{M^2}{2}+\frac{N^q}{2 }\sum\limits_{k=1}^{\infty}\frac{(-1)^{k}}{k}\(i\sqrt{\frac{2\l}{N^q}}\;M\)^{k}\right.\right.\nn\\
&\left.\left.+\frac{N\s^2}{2\;q!}\(1+\sum\limits_{k=1}^{\infty}(-1)^{k}\(i\sqrt{\frac{2\l}{N^q}}\;M\)^{k}\)\int \int d\t d\t'\sum\limits_{\a,\b}^{n}\(\frac{1}{N}\sum\limits_{i=1}^{N}\psi^{(\a)}_{i}(\t)\psi^{(\b)}_{i}(\t')\)^{q}\right)\right].\label{ZnMavg}
\eal

Considering the small $\lambda$ limit and keeping everything to the order $\l^{1/2}$ in the above expression for $\VEV{Z^{n}(J)}_{J}$, one arrives at the following in the large N limit
\bal
\VEV{Z^{n}(J)}_{J}&=  \int \cD\psi\;dM\;\exp\left[\left(-\frac{1}{2}\int d\t\sum\limits_{\a=1}^{n}\sum\limits_{i=1}^{N}\psi^{(\a)}_{i}\pa_{\t}\psi^{(\a)}_{i}-\frac{M^2}{2}-i\;\frac{N^q}{2}\sqrt{\frac{2\l}{N^q}}M\right.\right.\nn\\
&\left.\left.+N\frac{2^{q-2} \cJ^2}{q^2}\(1-i\sqrt{\frac{2\l}{N^q}}\;M\)\int \int d\t d\t'\sum\limits_{\a,\b}^{n}\(\frac{1}{N}\sum\limits_{i=1}^{N}\psi^{(\a)}_{i}(\t)\psi^{(\b)}_{i}(\t')\)^{q}\right)\right],\label{ZnMavg0}
\eal
where in the above expression we have used \eqref{varSYK} to replaced $\s$ in terms of $\cJ$. One can integrate out the auxiliary field M in \eqref{ZnMavg0} to obtain
\bal
\VEV{Z^{n}(J)}_{J}&=  \int \cD\psi\;\exp\left[\left(-\frac{1}{2}\int d\t\sum\limits_{\a=1}^{n}\sum\limits_{i=1}^{N}\psi^{(\a)}_{i}\pa_{\t}\psi^{(\a)}_{i}\right.\right.\nn\\
&\left.\left.+N\frac{2^{q-2} \cJ^2}{q^2}\(1-\l\)\int \int d\t d\t'\sum\limits_{\a,\b}^{n}\(\frac{1}{N}\sum\limits_{i=1}^{N}\psi^{(\a)}_{i}(\t)\psi^{(\b)}_{i}(\t')\)^{q}\right.\right.\nn\\
&\left.\left.-\frac{\l}{N^{q-2}}\frac{4^{q-2}\cJ^4}{q^4}\(\int \int d\t d\t'\sum\limits_{\a,\b}^{n}\(\frac{1}{N}\sum\limits_{i=1}^{N}\psi^{(\a)}_{i}(\t)\psi^{(\b)}_{i}(\t')\)^{q}\)\right.\right.\nn\\
&\left.\left.\(\int \int d\bar{\t} d\bar{\t}'\sum\limits_{\m,\n}^{n}\(\frac{1}{N}\sum\limits_{i=1}^{N}\psi^{(\m)}_{i}(\bar{\t})\psi^{(\n)}_{i}(\bar{\t}')\)^{q}\)\right)\right].\label{Znpsiavg}
\eal

\paragraph{Replica partition function in finite $\l$ and Large N approximation:} In the above paragraph we showed that one can approximate the replica partition function in the small $\l$ and large N limit. However, it is also possible to consider a finite $\l$  and a large N expansion. To see this we begin by considering $\VEV{Z^{n}(J)}_{J}$ in \eqref{ZnJavg} and expand it in Large N limit while keeping $\l$ fixed as
\bal
\VEV{Z^{n}(J)}_{J}=  & \(2\pi N^{1-q}\s^2\)^{\frac{1}{2}N^{q}}\int \cD\psi\;dM\;\exp\left[\left(-\frac{1}{2}\int d\t \sum\limits_{\a=1}^{n}\sum\limits_{i=1}^{N}\psi^{(\a)}_{i}\pa_{\t}\psi^{(\a)}_{i}\right.\right.\nn\\
&\left.\left.-\frac{M^2}{2}-i M \sqrt{\frac{\l N^q }{2 }}-\frac{\l}{2} M^2\right.\right.\nn\\
&\left.\left.+\frac{i^q N\s^2}{2~q!}\(1-i M \sqrt{\frac{2\l }{N^q}}\)\int \int d\t d\t'\sum\limits_{\a,\b}^{n}\(\frac{1}{N}\sum\limits_{i=1}^{N}\psi^{(\a)}_{i}(\t)\psi^{(\b)}_{i}(\t')\)^{q}+\cO(N^{-3q/2})\right)\right].\label{ZnJavgNe}
\eal

Once again integrating out the  auxiliary field M from the above action, one is left with the following
\bal
\VEV{Z^{n}(J)}_{J}&=  \int \cD\psi\;\exp\left[\left(-\frac{1}{2}\int d\t\sum\limits_{\a=1}^{n}\sum\limits_{i=1}^{N}\psi^{(\a)}_{i}\pa_{\t}\psi^{(\a)}_{i}\right.\right.\nn\\
&\left.\left.+N\frac{2^{q-2} \cJ^2}{q^2\(1+\l\)}\int \int d\t d\t'\sum\limits_{\a,\b}^{n}\(\frac{1}{N}\sum\limits_{i=1}^{N}\psi^{(\a)}_{i}(\t)\psi^{(\b)}_{i}(\t')\)^{q}\right.\right.\nn\\
&\left.\left.-\frac{\l}{N^{q-2}\(1+\l\)}\frac{4^{q-2}\cJ^4}{q^4}\(\int \int d\t d\t'\sum\limits_{\a,\b}^{n}\(\frac{1}{N}\sum\limits_{i=1}^{N}\psi^{(\a)}_{i}(\t)\psi^{(\b)}_{i}(\t')\)^{q}\)\right.\right.\nn\\
&\left.\left.\(\int \int d\bar{\t} d\bar{\t}'\sum\limits_{\m,\n}^{n}\(\frac{1}{N}\sum\limits_{i=1}^{N}\psi^{(\m)}_{i}(\bar{\t})\psi^{(\n)}_{i}(\bar{\t}')\)^{q}\)\right)\right].\label{Znpsiavg1}
\eal
where in the above expression for $\VEV{Z^{n}(J)}_{J}$ we have used  \eqref{varSYK} to replace $\s$ in terms of $\cJ$ as before. It is instructive to note that on taking small $\l$ limit in \eqref{Znpsiavg1} one goes back to \eqref{Znpsiavg} derived before in this section. In the rest of the article we will work with the replica partition function \eqref{Znpsiavg1}, and consider the effect of the finite coupling $\l$ on the properties of the SYK model. Having obtained the disorder averaged value of the replica partition function for the SYK model with non-Gaussian disorder in the large N limit, we will integrate out the fermions to get a bilocal effective action in the next subsection.

\subsection{Effective action in the Large N limit}
In this section we present the large N effective action $(I_{eff})$ for the SYK model which involves integrating out the fermions in \eqref{Znpsiavg1} through the introduction of bilocal fields $\S(\t,\t')$ and $G(\t,\t')$ which are related to the Majorana fermionic fields of the SYK model. In order to achieve this one has to first define the bilocal field
\bal
\tilde{G}_{\a,\b}(\t,\t')=\frac{1}{N}\sum\limits_{i=1}^{N}\psi^{(\a)}_{i}(\t)\psi^{(\b)}_{i}(\t'),\label{Gconstarint}
\eal
and then introduce the Lagrange multiplier field $\tilde{\S}_{\a,\b}$ which enforces the constraint given above. To enforce the constraint \eqref{Gconstarint} we include in the action \eqref{Znpsiavg1}, the following identity 
\bal
\int \cD\tilde{\S}\;\cD \tilde{G}\;\exp\left[-\frac{N}{2}\sum\limits_{\a,\b}^{n}\int\int d\t d\t'\;\tilde{\S}_{\a,\b}(\t,\t')\(\tilde{G}_{\a,\b}(\t,\t')-\frac{1}{N}\sum\limits_{i=1}^{N}\psi^{(\a)}_{i}(\t)\psi^{(\b)}_{i}(\t')\)\right]=1,\label{GSconstraint}
\eal
which leads to the following expression for the replica partition function
\bal
\VEV{Z^{n}(J)}_{J}=&\int\cD\tilde{\S}\;\cD\tilde{G}\left(\int \cD\psi \exp\left[\left(-\frac{1}{2\;N}\int d\t \sum\limits_{\a=1}^{n}\sum\limits_{i=1}^{N}\psi^{(\a)}_{i}\pa_{\t}\psi^{(\a)}_{i}\right.\right.\right.\nn\\
&\left.\left.\left.+\frac{2^{q-2} \cJ^2}{q^2\(1+\l\)}\int \int d\t d\t'\sum\limits_{\a,\b}^{n}\tilde{G}_{\a,\b}(\t,\t')^{q}\right.\right.\right.\nn\\
&\left.\left.\left.-\frac{\l}{N^{q-1}\(1+\l\)}\frac{4^{q-2}\cJ^4}{q^4}\(\int \int d\bar{\t} d\bar{\t}'\sum\limits_{\m,\n}^{n}\tilde{G}_{\m,\n}(\bar{\t},\bar{\t}')^{q}\)\(\int \int d\t d\t'\sum\limits_{\a,\b}^{n}\tilde{G}_{\a,\b}(\t,\t')^{q}\)\right.\right.\right.\nn\\
&\left.\left.\left.-\frac{1}{2}\int\int d\t d\t' \sum\limits_{\a,\b}^{n}\tilde{\S}_{\a,\b}(\t,\t')\(\tilde{G}_{\a,\b}(\t,\t')-\frac{1}{N}\sum\limits_{i=1}^{N}\psi^{(\a)}_{i}(\t)\psi^{(\b)}_{i}(\t')\)\right)\right]\right)^{N},
\eal
such that on introducing the rescalings $\cJ\to n^{-\frac{1}{2q}} \cJ$ and $N\to n^{\frac{1}{q}} N$ in the  replica partition function above, it reduces to the following \footnote{The authors in \cite{Krajewski:2018lom} have not considered such a rescaling and it is not clear how they obtain a replica diagonal solution to the effective action in their case.} 
\bal
\VEV{Z^{n}(J)}_{J}=& \int\cD\tilde{\S}\;\cD\tilde{G}\left(\int \cD\psi \exp\left[n\left(-\frac{1}{2n}\int\int d\t d\t'\sum\limits_{\a\b}^{n}\psi^{(\a)}(\t)\(\d_{\a\b}\d(\t-\t')\pa_{\t}-\tilde{\S}_{\a,\b}(\t,\t')\)\psi^{(\b)}(\t')\right.\right.\right.\nn\\
&\left.\left.\left.-\frac{1}{2n}\int\int d\t d\t' \sum\limits_{\a,\b}^{n}\tilde{\S}_{\a,\b}(\t,\t')\tilde{G}_{\a,\b}(\t,\t')+\frac{2^{q-2} \cJ^2}{n~q^2\(1+\l\)}\int \int d\t d\t'\sum\limits_{\a,\b}^{n}\tilde{G}_{\a,\b}(\t,\t')^{q}\right.\right.\right.\nn\\
&\left.\left.\left.-\frac{\l}{N^{q-1}\(1+\l\)}\frac{4^{q-2}\cJ^4}{n^2q^4}\(\int \int d\bar{\t} d\bar{\t}'\sum\limits_{\m,\n}^{n}\tilde{G}_{\m,\n}(\bar{\t},\bar{\t}')^{q}\)\(\int \int d\t d\t'\sum\limits_{\a,\b}^{n}\tilde{G}_{\a,\b}(\t,\t')^{q}\)\right)\right]\right)^{N}.\label{ZnSGeff}
\eal

In the large N limit the integrals over $\cD\tilde{\S}\;\cD\tilde{G}$ can be performed in \eqref{ZnSGeff} by finding a saddle point such that
\bal
-\ln\(\VEV{Z^{n}(J)}_{J}\)=&-\ln\(\int \cD\tilde{\S}\;\cD\tilde{G}\; \exp\(-I^{(n)}[\tilde{\S},\tilde{G}]\)\),\nn\\
\simeq & \;\(\min\limits_{\S} \max\limits_{G} I^{(n)}[\S,G]\),\label{mimmaxI}
\eal
where
\bal
I^{(n)}[\S,G]&=n~N\left(-\frac{1}{2n} \ln \det\(\d_{\a\b}\d(\t-\t')\pa_{\t}-\S_{\a,\b}(\t,\t')\)\right.\nn\\
&\left.+\frac{1}{2n}\int\int d\t d\t' \sum\limits_{\a,\b}^{n}\S_{\a,\b}(\t,\t')G_{\a,\b}(\t,\t')-\frac{2^{q-2} \cJ^2}{n~q^2\(1+\l\)}\int \int d\t d\t'\sum\limits_{\a,\b}^{n}G_{\a,\b}(\t,\t')^{q}\right.\nn\\
&\left.+\frac{\l}{N^{q-1}\(1+\l\)}\frac{4^{q-2}\cJ^4}{n^2q^4}\(\int \int d\bar{\t} d\bar{\t}'\sum\limits_{\m,\n}^{n}G_{\m,\n}(\bar{\t},\bar{\t}')^{q}\)\(\int \int d\t d\t'\sum\limits_{\a,\b}^{n}G_{\a,\b}(\t,\t')^{q}\)\right),\label{Ineff0}
\eal
such that the bilocal fields $\S_{\a,\b}(\t,\t')$ and $G_{\a,\b}(\t,\t')$ now denote the saddle point solutions to the Schwinger Dyson equations 
\bal
\S_{\a,\b}(\t,\t')&=\frac{2^{q-1} \cJ^2}{q\(1+\l\)}G_{\a,\b}(\t,\t')^{q-1}\nn\\
&-\frac{\l}{N^{q-1}\(1+\l\)}\frac{4^{q-1}\cJ^4}{n~q^3}G_{\a,\b}(\t,\t')^{q-1}\(\sum\limits_{\m,\n}^{n}\int\int d\bar{\t} d\bar{\t}'G_{\m,\n}(\bar{\t},\bar{\t}')^{q}\),\label{sdG}\\
G_{\a,\b}(\t,\t')&=\(\d_{\a\b}\d(\t-\t')\pa_{\t}-\S_{\a,\b}(\t,\t')\)^{-1}.\label{sdS}
\eal

In order to solve the above equations of motion, one may consider the usual replica diagonal ansatz for which $\S_{\a,\b}(\t,\t')=\d_{\a\b}\S(\t,\t')$ and $G_{\a,\b}(\t,\t')=\d_{\a\b}G(\t,\t')$, where $\S(\t,\t')$ and $G(\t,\t')$ denote the replica diagonal saddle point solutions. Assuming such an ansatz it is obvious to  see that
\bal
I^{(n)}[\S,G]&=n~N\left(-\frac{1}{2} \ln \det\(\d(\t-\t')\pa_{\t}-\S(\t,\t')\)\right.\nn\\
&\left.+\frac{1}{2}\int\int d\t d\t' \S(\t,\t')G(\t,\t')-\frac{2^{q-2} \cJ^2}{q^2\(1+\l\)}\int \int d\t d\t'G(\t,\t')^{q}\right.\nn\\
&\left.+\frac{\l}{N^{q-1}\(1+\l\)}\frac{4^{q-2}\cJ^4}{q^4}\(\int \int d\bar{\t} d\bar{\t}'G(\bar{\t},\bar{\t}')^{q}\)\(\int \int d\t d\t'G(\t,\t')^{q}\)\right),\label{Ineff1}
\eal
Using \eqref{reptrick}, it is now trivial to take the replica limit
\bal
-\VEV{\ln Z(J)}_{J}=-\lim_{n\to 0}\frac{\VEV{e^{(-n\;I_{eff}(J))}}-1}{n}\simeq I_{eff}(J)
\eal
where 
\bal
\frac{I_{eff}(J,\l)}{N}&=-\frac{1}{2} \Tr\ln \(\d(\t-\t')\pa_{\t}-\S(\t,\t')\)+\frac{1}{2}\int\int d\t d\t'\S(\t,\t')G(\t,\t')\nn\\
&-\frac{2^{q-2}\cJ^2}{ q^2\(1+\l\)}\int \int d\t d\t'G(\t,\t')^{q}\nn\\
 &+\frac{\l}{N^{q-1}\(1+\l\)}\frac{ 4^{q-2}\cJ^4}{q^4}\(\int \int d\t d\t'G(\t,\t')^{q}\)\(\int \int d\bar{\t}d\bar{\t}'G(\bar{\t},\bar{\t}')^{q}\),\label{IneffFin}
\eal   
such that the Schwinger Dyson equations in \eqref{sdG} and \eqref{sdS} now become 
\bal
\S(\t,\t')&=\frac{2^{q-1}\cJ^2}{ q\(1+\l\)}G(\t,\t')^{q-1}-\frac{\l}{N^{q-1}\(1+\l\)}\;\frac{ 4^{q-1}\cJ^4}{q^3}G(\t,\t')^{q-1}\(\int\int d\bar{\t} d\bar{\t}'G(\bar{\t},\bar{\t}')^{q}\),\label{sdGF}\\
G(\t,\t')&=\(\d(\t-\t')\pa_{\t}-\S(\t,\t')\)^{-1},\label{sdSF}
\eal
for the effective action $I_{eff}$ given above. To this end, we would like to emphasize that we have kept the subdominant $\cJ^4$ term in the effective action \eqref{IneffFin} and consequently in the Schwinger Dyson equations, in order to study the leading order perturbative corrections to the SYK model with non-Gaussian disorder in the Large $N$ and finite $\l$ limit. As one can see, the non-Gaussian disorder has two effects. First it renormalizes the bare propagator $G$ such that $\mathcal{J}^2\rightarrow \mathcal{J}^2/(1+\lambda)$ and second, it induces a non-local term in the action suppressed by powers of $N^{q-1}$ in the large N limit.

\subsection{Partition and Correlations functions}
In this subsection, we outline the prescription for calculating the correlation functions in the SYK model with non-Gaussian disorder. For this, we will first focus on computing the thermal partition function using a particular method and then show how this method can be used to evaluate the general n-point correlation function. 

\paragraph{Thermal Partition function} 
 In order to evaluate the thermal partition function for the non-Gaussian SYK model \cite{Gross:2019uxi}, we start with the large $N$ effective action for the SYK model with Gaussian disorder, i.e. we set $\l=0$ in \eqref{IneffFin} to obtain the following
\bal
\frac{I_{eff}(J)}{N}&=-\frac{1}{2} \Tr\ln \(\d(\t-\t')\pa_{\t}-\S(\t,\t')\)+\frac{1}{2}\int d\t\left(\int  d\t'\S(\t,\t')G(\t,\t')\right.\nn\\
&\left.-\frac{2^{q-1}\cJ^2}{ q^2}\int d\t'G(\t,\t')^{q}\right) ,\label{SYKActionE}
\eal  
where one can identify the interaction Hamiltonian for the SYK model with Gaussian disorder as
\bal
H_0=- N\left(\frac{2^{q-2}\cJ^2}{ q^2}\int d\t'G(\t,\t')^{q} \right).\label{intH}
\eal

Given the form of the interaction Hamiltonian for the SYK model above, we write the interaction Hamiltonian for the case of non-Gaussian disorder in \eqref{IneffFin} as
\bal
H=- N\left(\frac{2^{q-2}\cJ^2}{ q^2(1+\l)}\int d\t'G(\t,\t')^{q} \right),\label{intHB}
\eal
where it can be observed that the deformation parameter only rescales the coupling $\cJ$ of the SYK model. Moreover, the total interaction term in \eqref{IneffFin} can also be written in terms of $H$ defined in \eqref{intHB} as
\bal
f(\l,H)&=-\frac{2^{q-2}\cJ^2}{ q^2\(1+\l\)} \int d\t'G(\t,\t')^{q}
 +\frac{\l \b}{N^{q-1}\(1+\l\)}\frac{ 4^{q-2}\cJ^4}{q^4}\(\int  d\t'G(\t,\t')^{q}\)\(\int d\bar{\t}'G(\bar{\t},\bar{\t}')^{q}\),\nn\\
 &=\frac{H}{N}+\frac{\l\b}{N^{q-1}}\(1+\l\) \(\frac{H}{N}\)^2,\label{deffunc}
\eal
as $G(\t,\t')$ is considered to be a function of the difference of its arguments which is why there is a factor of $\beta$ present in the second term of the above expression. Now using a Hubbard-Stratonovich transformation one can write the thermal partition function for the SYK model with non-Gaussian disorder as 
\bal
\cZ(\l,\b)&=\Tr{\(e^{-\b f(\l,H)}\)}=\frac{1}{\sqrt{2\pi}}\int\limits_{-\infty}^{\infty} d\tilde{M}\int\limits_{-\infty}^{\infty}dE\;e^{-\frac{\tilde{M}^2}{2}-\b \(1+i\sqrt{\frac{2\l(1+\l)}{N^{q-1}}}\;\tilde{M}\)E}\;\r(E),\label{defZ}
\eal

It is known that in the low temperature regime $(\b\to \infty)$, the dynamics of the SYK model is described by a Schwarzian theory \cite{Maldacena:2016hyu,Gross:2019uxi}. In this regime, we understand the bulk dual for the SYK model is described by JT gravity in $AdS_2$ with Dirichlet conditions for the dilaton and the metric at the AdS boundary \cite{Maldacena:2016hyu,Maldacena:2016upp}. The partition function and the density of states for the SYK model in the small temperature regime can be given as
\bal
Z^{syk}_{0}(\b)=\frac{1}{4\sqrt{\pi\b^3}}e^{\frac{\pi^2}{\b}},\quad \r_{0}(E)=\frac{\sinh{\(2\pi\sqrt{E}\)}}{4\pi^2},\label{UdefZ}
\eal

Now on substituting the density of the states for the SYK model with Gaussian disorder, $\r(E)=\r_0(E)$ defined in \eqref{UdefZ} in the relation \eqref{defZ}, one can evaluate the partition function for the non-Gaussian SYK model as
\bal
\cZ(\l,\b)=&\int\limits_{-\infty}^{\infty} d\tilde{M}\;\frac{1}{\sqrt{32\pi^2\b^3}}\;\exp{\[-\frac{\tilde{M}^2}{2}+\frac{\pi^2}{\b\(1+i\sqrt{\frac{2\l(1+\l)}{N^{q-1}}}\;\tilde{M}\)}-\log{\(1+i\sqrt{\frac{2\l(1+\l)}{N^{q-1}}}\;\tilde{M}\)}\]},\nn\\
\simeq & \int\limits_{-\infty}^{\infty} d\tilde{M}\;\frac{1}{\sqrt{32\pi^2\b^3}}\;\left(1-i\(1+\frac{\pi^2}{\b}\)\sqrt{\frac{2\l(1+\l)}{N^{q-1}}}\;\tilde{M}-\frac{\left(2 \beta ^2+4 \pi ^2 \beta +\pi ^4\right) \l (1+\l) }{\beta ^2N^{q-1}}\tilde{M}^2\right.\nn\\
&\left.+\cO\(N^{-\frac{3(q-1)}{2}}\)\right)\exp{\[-\frac{\tilde{M}^2}{2}+\frac{\pi^2}{\b}\]},\nn\\
&=\frac{1}{4\sqrt{\pi\b^3}}\(1-\frac{\left(2 \beta ^2+4 \pi ^2 \beta +\pi ^4\right) \l (1+\l) }{\beta ^2N^{q-1}}\)\exp{\[\frac{\pi ^2}{\beta }\]},\label{modZ}
\eal
where in the second line we have performed a large N expansion to approximate the partition function. In order to determine the density of states for the non-Gaussian SYK model we first write the partition function in \eqref{defZ} as
\bal
\cZ(\l,\b)=\frac{1}{\sqrt{2\pi}}\int\limits_{-\infty}^{\infty} d\tilde{M}\int\limits_{-\infty}^{\infty}d\bar{E}\;\frac{e^{-\frac{\tilde{M}^2}{2}-\b \bar{E}}}{\(1+i\sqrt{\frac{2\l(1+\l)}{N^{q-1}}}\;\tilde{M}\)}\;\r\(\frac{\bar{E}}{\(1+i\sqrt{\frac{2\l(1+\l)}{N^{q-1}}}\;\tilde{M}\)}\),\label{defRZ}
\eal
where in deriving the second line we have considered the transformation
\bal
E\to \frac{\bar{E}}{\(1+i\sqrt{\frac{2\l(1+\l)}{N^{q-1}}}\;\tilde{M}\)}.\label{EtobE}
\eal

As $\bar{E}$ is just an integration variable, one can replace it by $E$ in \eqref{defZ} which further leads us to write
\bal
\cZ(\l,\b)&=\frac{1}{\sqrt{2\pi}}\int\limits_{-\infty}^{\infty} d\tilde{M}\int\limits_{-\infty}^{\infty}dE\;\frac{e^{-\frac{\tilde{M}^2}{2}-\b E}}{\(1+i\sqrt{\frac{2\l(1+\l)}{N^{q-1}}}\;\tilde{M}\)}\;\r\(\frac{E}{\(1+i\sqrt{\frac{2\l(1+\l)}{N^{q-1}}}\;\tilde{M}\)}\),\nn\\
&=\int\limits_{-\infty}^{\infty}dE\;e^{-\b E}\r_{\l}(E),\label{defZ1}
\eal
where the modified density of states $\r_{\l}(E)$ is now given by
\bal
\r_{\l}(E)=\frac{1}{\sqrt{2\pi}}\int\limits_{-\infty}^{\infty} d\tilde{M}\;e^{-\frac{\tilde{M}^2}{2}-\log{\(1+i\sqrt{\frac{2\l(1+\l)}{N^{q-1}}}\;\tilde{M}\)}}\;\r\(\frac{E}{\(1+i\sqrt{\frac{2\l(1+\l)}{N^{q-1}}}\;\tilde{M}\)}\).\label{defr}
\eal

\begin{figure}[h]
    \centering
    \captionsetup{width=0.85\linewidth}
    \includegraphics[width=0.5\textwidth]{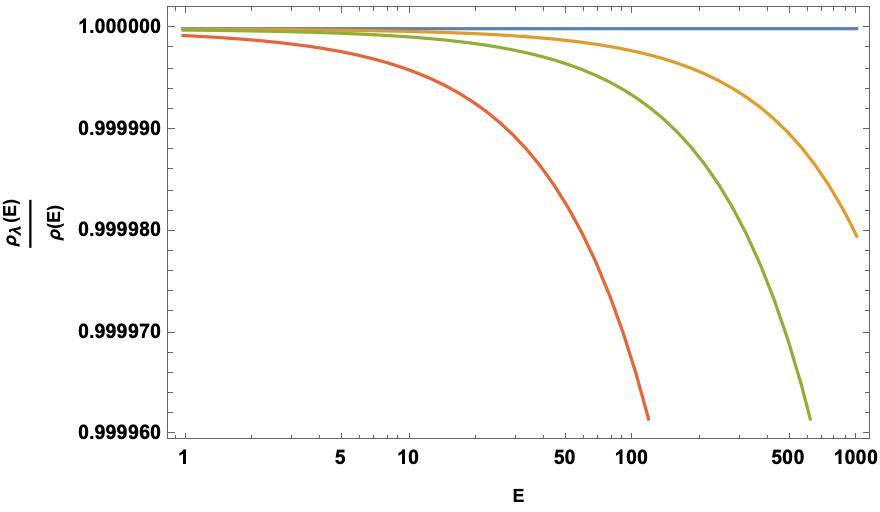}
    \caption{LogLog plot of $\frac{\r_{\l}(E)}{\r_0(E)}$ with respect to the energy E for different values of $\l=\left\{0,1,2,5\right\}$ represented by the blue, orange, green, and red curves respectively. We have fixed the parameters $N=1000$ and $q=4$ for all values of $\l$.}
    \label{fig:MDplot}
\end{figure}

Using the functional form of the density of states $\r(E)=\r_0(E)$ defined in \eqref{UdefZ}, the modified density of states in \eqref{defr} can be evaluated as
\bal
\r_{\l}(E)&=\frac{1}{\sqrt{32\pi^5}}\int\limits_{-\infty}^{\infty} d\tilde{M}\;e^{-\frac{\tilde{M}^2}{2}-\log{\(1+i\sqrt{\frac{2\l(1+\l)}{N^{q-1}}}\;\tilde{M}\)}}\;\sinh\(\frac{2\pi \sqrt{E}}{\(1+i\sqrt{\frac{2\l(1+\l)}{N^{q-1}}}\;\tilde{M}\)^{1/2}}\),\nn\\
&\simeq \frac{1}{\sqrt{32\pi^5}}\int\limits_{-\infty}^{\infty} d\tilde{M}\;\exp{\[-\frac{\tilde{M}^2}{2}\]}\left(1-i\(\sinh \left(2 \pi  \sqrt{E}\right)+\pi  \sqrt{E} \cosh \left(2 \pi  \sqrt{E}\right)\)\sqrt{\frac{2\l(1+\l)}{N^{q-1}}}\;\tilde{M}\right.\nn\\
&\left.-\frac{\l(1+\l)}{2N^{q-1}}\(2 \left(\pi ^2 E+2\right) \sinh \left(2 \pi  \sqrt{E}\right)+7 \pi  \sqrt{E} \cosh \left(2 \pi  \sqrt{E}\right)\)\tilde{M}^2+\cO\(N^{-\frac{3(q-1)}{2}}\)\right),\nn\\
&=\frac{\sinh{\(2\pi\sqrt{E}\)}}{4\pi^2}\(1-\left(\left(2+\pi ^2 E\right) \frac{\l(1+\l)}{N^{q-1}}\right)- \frac{7\l(1+\l)}{2N^{q-1}} \pi  \sqrt{E} \coth \left(2 \pi  \sqrt{E}\right)\),\label{modr}
\eal
which shows that the non-Gaussian model starts with a similar density of states at low energy, but the density decays exponentially at higher energies compared to the Gaussian SYK as shown in the figure \eqref{fig:MDplot}. 
The procedure described above for computing the low-temperature partition function for the non-Gaussian SYK model shows that one has to introduce an auxiliary field through a Hubbard-Stratonovich transformation. This transformation relates the physical quantities in the Gaussian SYK model to the corresponding physical quantities in the non-Gaussian SYK model. As we will elucidate later, implementing a Hubbard-Stratonovich transformation will be our method for obtaining the correlation functions and other relevant physical quantities pertaining to the non-Gaussian SYK model.

\paragraph{Correlation Functions:} Before we focus on the computation of the correlation functions for the non-Gaussian SYK model it is important to understand how they are evaluated for the case of Gaussian disorder. For this, we first consider an operator $\cO$ made up of the product of Majorana fermion bilinears for the usual SYK model. The finite temperature expectation value of such an operator can be given as
\bal
\VEV{\cO}_{\b}&=\int\limits_{0}^{\infty} dE\;e^{-\b E}\cO(E),\quad \cO(E)=\VEV{ E\abs{\cO} E},\label{ThO1pSYKG}
\eal
where the vacuum expectation value of the operator $\cO$ is simply given by $\la 0\abs{\cO}0\ra$. We have defined the finite temperature expectation value of the operator in the sense that when $\cO$ is an identity operator its expectation value is just the thermal partition function.

The finite temperature expectation value of the operator $\cO$ can now be evaluated for the SYK with the non-Gaussian disorder as
\bal
\VEV{\cO}_{\l,\b}&=\int\limits_{0}^{\infty} dE\;e^{-\b f(\l,E)}\cO(E),\nn\\
&=\frac{1}{\sqrt{2\pi}}\int\limits_{-\infty}^{\infty} d\tilde{M}\int\limits_{0}^{\infty}dE\;e^{-\frac{\tilde{M}^2}{2}-\b \(1+i\sqrt{\frac{2\l(1+\l)}{N^{q-1}}}\;\tilde{M}\)E} \cO(E),\label{ThO1p}
\eal
where the function $f(\l, H)$ is defined in \eqref{deffunc} and in writing the last line in the above expression we have used a Hubbard-Stratonovich transformation similar to the case of the thermal partition discussed in \eqref{defZ} before. Here it is to be noted that the thermal expectation value of the operator $\VEV{\cO}_{\l,\b}$ with the subscript $\l$ denotes its value for the non-Gaussian SYK model. From \eqref{ThO1p} it is trivial to see that as $\l\to 0$, the expectation value of the operator goes to the one that corresponds to the case of the Gaussian SYK model in \eqref{ThO1pSYKG}. Now in order to relate the expectation values of the operator $\cO$ in the two SYK models with different disorder averaging we consider the following transformation \eqref{EtobE} which reduces \eqref{ThO1p} to the following
\bal
\VEV{\cO}_{\l,\b}&=\frac{1}{\sqrt{2\pi}}\int\limits_{-\infty}^{\infty} d\tilde{M}\int\limits_{0}^{\infty}d\bar{E}\;\frac{e^{-\frac{\tilde{M}^2}{2}-\b \bar{E}}}{\(1+i\sqrt{\frac{2\l(1+\l)}{N^{q-1}}}\;\tilde{M}\)} \cO\(\frac{\bar{E}}{\(1+i\sqrt{\frac{2\l(1+\l)}{N^{q-1}}}\;\tilde{M}\)}\),\label{ThO1pSYKNG}
\eal
which can be further written as
\bal
\VEV{\cO}_{\l,\b}&=\int\limits_{0}^{\infty}dE\;e^{-\b E}\;\cO(E)_{\l},\nn\\
\cO(E)_{\l}&=\frac{1}{\sqrt{2\pi}}\int\limits_{-\infty}^{\infty} d\tilde{M}\;\frac{e^{-\frac{\tilde{M}^2}{2}}}{\(1+i\sqrt{\frac{2\l(1+\l)}{N^{q-1}}}\;\tilde{M}\)} \cO\(\frac{E}{\(1+i\sqrt{\frac{2\l(1+\l)}{N^{q-1}}}\;\tilde{M}\)}\),\label{ThO1pSYKNGE}
\eal
where we have removed the bar over the energy as it is only an integration variable. Here it is to be noted that $\cO(E)_{\l}$ stands for the expectation value of the operator for the energy eigenstate $\ket{E}$ in the case of the non-Gaussian SYK model. This is related to $\cO(E)$ through the relation \eqref{ThO1pSYKNGE}, where $\cO(E)$ is the expectation value of the operator for the energy eigenstate $\ket{E}$ in the case of the Gaussian SYK model. Thus, the relation in \eqref{ThO1pSYKNGE} gives us a dictionary to relate the thermal correlation functions in the two SYK models with different disorder averaging.

Now in order to demonstrate the computation of a general n-point correlation function of the operator $\cO$, we begin by considering the simple case of the vacuum one-point function of the same operator. It can be given as
\bal
\VEV{\cO(\t)}_{\l,0}=\frac{1}{\sqrt{2\pi}}\int\limits_{-\infty}^{\infty} d\tilde{M}\int\limits_{0}^{\infty}dE\;e^{-\frac{\tilde{M}^2}{2}-\t \(1+i\sqrt{\frac{2\l(1+\l)}{N^{q-1}}}\;\tilde{M}\)E} \cO(E),\label{ZTO1p}
\eal
where we have replaced $\b$ by $\t$ in \eqref{ThO1p} assuming that $\t>0$ and the ground state energy is zero. In a similar way, the n-point vacuum correlator for the non-Gaussian SYK model can be obtained 
\bal
&\VEV{\cO(\t_1)\cO(\t_2)\cdots\cO(\t_{n-1})\cO(0)}_{\l,0}= \frac{1}{\sqrt{2\pi}}\int\limits_{-\infty}^{\infty} d\tilde{M} \;e^{-\frac{\tilde{M}^2}{2}} \int\limits_{0}^{\infty}\prod\limits_{i}^{n-1}dE_{i}\left(\VEV{0\abs{\cO(\t_1)}E_1}\VEV{E_1\abs{\cO(\t_2)}E_2}\right.\nn\\
&\left.\VEV{E_2\abs{\cO(\t_3)}E_3}\cdots\VEV{E_{n-1}\abs{\cO(0)}0}\right)\exp{\[-\(1+i\sqrt{\frac{2\l(1+\l)}{N^{q-1}}}\;\tilde{M}\)\(\sum\limits_{i=1}^{n-2}(\t_i-\t_{i+1})E_{i}+\t_{n-1}E_{n-1}\)\]},\label{ZTOnp}
\eal
whereas the finite temperature n-point correlation function can be given as
\bal
&\VEV{\cO(\t_1)\cO(\t_2)\cdots\cO(\t_{n-1})\cO(\t_n)}_{\l,\b}= \frac{1}{\sqrt{2\pi}}\int\limits_{-\infty}^{\infty} d\tilde{M} \;e^{-\frac{\tilde{M}^2}{2}} \int\limits_{0}^{\infty}\prod\limits_{i}^{n-1}dE_{i}\left(\VEV{E_1\abs{\cO(\t_1)}E_2}\VEV{E_2\abs{\cO(\t_2)}E_3}\right.\nn\\
&\left.\VEV{E_3\abs{\cO(\t_3)}E_4}\cdots\VEV{E_{n-1}\abs{\cO(\t_n)}E_n}\right)\exp{\[-\(1+i\sqrt{\frac{2\l(1+\l)}{N^{q-1}}}\;\tilde{M}\)\(\sum\limits_{i=1}^{n}\b_{i}E_{i}\)\]},\label{ThOnp}
\eal
where we identified, $\t_i -\t_{i+1}=\b_i$ for $i=\{1,2,\cdots,n-1\}$, $\t_1=\b_1$ and $\t_n=\b_n$. To this end, it should be noted that the expectation values and the energy eigenstates in the right-hand side of the expressions \eqref{ZTO1p} to \eqref{ThOnp} correspond to the Gaussian SYK model.

Having given the general prescription for evaluating the n-point correlation function we now consider correlators in one-dimensional conformal field theory as an example. For instance, the two-point zero temperature correlator in this case can be given as
\bal
\VEV{\cO(\t)\cO(0)}_{0,0}=\frac{1}{\t^{2\D}}=\int\limits_{0}^{\infty}dE\;e^{-\t E} \abs{\VEV{0\abs{\cO} E}}^2,\quad \abs{\VEV{0\abs{\cO} E}}^2=\frac{E^{2\D-1}}{\Gamma{(2\D)}}
\eal
where $\D$ is the scaling dimension of the operator. Plugging the value of $\abs{\VEV{0\abs{\cO} E}}^2$ from above in \eqref{ZTOnp} for $n=2$ and $E_1=E$, one gets the two-point function in the modified conformal field theory with non zero $\l$ as
\bal
\VEV{\cO(\t)\cO(0)}_{\l,0}&\simeq \frac{1}{\sqrt{2\pi}}\int\limits_{-\infty}^{\infty} d\tilde{M}\;\frac{\exp{\[-\frac{\tilde{M}^2}{2}\]}}{\tau ^{2 \Delta }(1+i\sqrt{\frac{2\l(1+\l)}{N^{q-1}}}\;\tilde{M})^{2\D}},\nn\\
&=\frac{1}{\t^{2\D}}\(1-2 \Delta  (2 \Delta +1) (\lambda +1) \frac{\l}{N^{q-1}}+\cO\(N^{-\frac{3(q-1)}{2}}\)\)
\eal
where we have obtained the above result by first integrating over the energy eigenstates and then integrating over the auxiliary variable $\tilde{M}$ in the large N approximation.

We can now use the above formulation to determine the correlation functions for the non-Gaussian SYK model. For this, we first consider the Schwinger-Dyson equations \cite{Maldacena:2016hyu} in frequency space for the Gaussian SYK model, 
\bal
G^{syk}(i\omega)=\frac{1}{-i \omega-\S^{syk}(i\omega)},\quad \S^{syk}(i\omega)=\frac{2^{q-1}\cJ^2}{q}G^{syk}(i\omega)^{q-1},\label{GSsyk}
\eal
which can be solved analytically for $q=2$ case giving the thermal two-point function $G^{syk}(i\omega)$ as
\bal
G^{syk}(i\omega)=- \frac{2}{i \om+ i  \sgn{(\om)}\sqrt{4\cJ^2+\om^2}},\quad \Sigma^{syk} (i\omega )=-\frac{i}{2}  \left(\omega+\sgn{(\om)}\sqrt{4 \cJ^2+\omega ^2} \right)\label{GsykW}
\eal
where $\omega=\frac{(2n+1)\pi}{\b}$ is the Matsubara frequency. The above two-point function $G^{syk}(i\omega)$  can also be written in terms of energy $E=i \omega$, as
\bal
G^{syk}(E)=- \(\frac{2}{E+ i \sqrt{4\cJ^2-E^2}}\),\label{GsykE}
\eal
such that the density of states can now be computed from the above two point function as
\bal
\r^{syk}(E)=\Im(G^{syk}(E))=\frac{1}{\cJ}\sqrt{1-\(\frac{E}{2\cJ}\)^2},\label{rhosyk}
\eal
which clearly follows the Wigner's Semi-circle law when plotted against the energy \cite{wigner_1951}. 

\begin{figure}[h]
    \centering
    \captionsetup{width=0.85\linewidth}
    \includegraphics[width=0.5\textwidth]{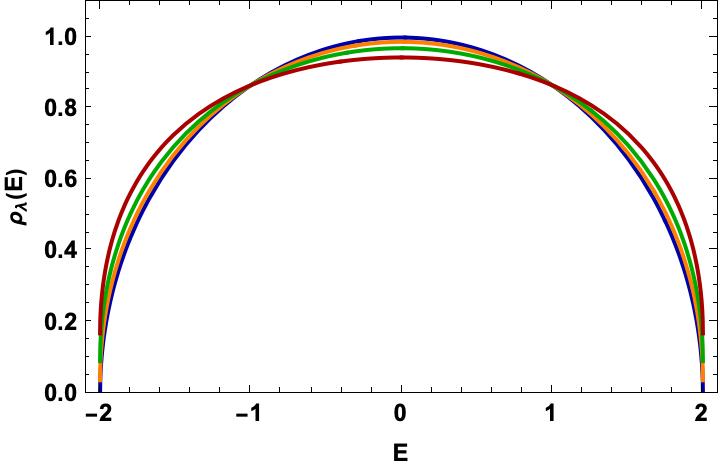}
    \caption{Plot of density of states $\r_{\l}(E)$ (for $q=2$ case) with respect to the energy E for different values of $\l=\left\{0,3,5,7\right\}$ represented by the blue, orange, green, and red curves respectively. We have fixed the parameters $N=1000$ and $\cJ=1$  for all values of $\l$.}
    \label{fig:q2MDplot}
\end{figure}

The thermal two-point function in \eqref{GsykE} can now be used to deduce the thermal two-point function for the non-Gaussian SYK model. For this, we first identify $G^{syk}(E)=\cO(E)$ and then on using the relation in \eqref{ThO1pSYKNGE} one can obtain the corresponding quantity in the case of non-Gaussian SYK as
\bal
G(E)_{\l}&=\frac{1}{\sqrt{2\pi}}\int\limits_{-\infty}^{\infty} d\tilde{M}\;\frac{e^{-\frac{\tilde{M}^2}{2}}}{\(1+i\sqrt{\frac{2\l(1+\l)}{N}}\;\tilde{M}\)} G^{syk}\(\frac{E}{\(1+i\sqrt{\frac{2\l(1+\l)}{N}}\;\tilde{M}\)}\),\nn\\
&=\frac{1}{\sqrt{2\pi}}\int\limits_{-\infty}^{\infty} d\tilde{M}\;\exp\left[-\frac{\tilde{M}^2}{2}-i\left(1+\frac{i E}{\sqrt{4 \cJ^2-E^2}}\right)\sqrt{\frac{2\l(1+\l)}{N}}\;\tilde{M}\right.\nn\\
&\left.~~~~~~~~~~~~~~~~~~~~~~~~~+\cO\(\frac{1}{N}\)\right]G^{syk}(E),\label{GIsykNGE}
\eal
such that after integrating over the auxiliary field $\tilde{M}$ one gets
\bal
G(E)_{\l}=- \(\frac{2}{E+ i \sqrt{4\cJ^2-E^2}}\)-\frac{2 \lambda  (1+\lambda)}{N}\(\frac{ \left(E-i \sqrt{4 \cJ^2-E^2}\right)^2}{\left(4 \cJ^2-E^2\right) \left(E+i \sqrt{4 \cJ^2-E^2}\right) }\)+\cO\left(\frac{1}{N^{2}}\right)\label{GsykNGE}
\eal
and the density of states corresponding to the above two-point function can now be given as 
\bal
\r_{\l}(E)=\frac{1}{\cJ}\sqrt{1-\(\frac{E}{2\cJ}\)^2}\(1-\frac{2\l (1+\l)}{N}\frac{\(\cJ^2-E^2\)}{\cJ^2 \sqrt{4 \cJ^2-E^2}}+\cO\left(\frac{1}{N^{2}}\right)\)
\eal
which shows the deviation of order $\cO\left(\frac{1}{N}\right)$ and higher from the semi-circle law obtained in \eqref{rhosyk} for the case of the Gaussian SYK model. This deviation from the semi-circle law for the density of states corresponding to $q=2$ non-Gaussian SYK model is depicted in figure \eqref{fig:q2MDplot} which shows that for fixed $\cJ$ and $N$, one gets a more squashed semi-circular behaviour for the density of states as the parameter $\l$ is increased.

As demonstrated for the $q=2$ case, if the usual Gaussian SYK n-point correlation functions are known in the frequency/energy representation then using the above procedure one can easily determine the n-point correlation functions for the non-Gaussian SYK model.

\section{Large q approximation and relation to Liouville field theory}

In order to study the thermodynamical properties of the SYK model one has to evaluate the free energy. It can be determined in the large N limit if one has the exact solution for the fields $\S(\t,\t')$ and $G(\t,\t')$ satisfying the Schwinger-Dyson equations described in the previous section. However, solving the Schwinger Dyson equations in \eqref{sdGF} and \eqref{sdSF} analytically for $\S(\t,\t')$ and $G(\t,\t')$ is a bit nontrivial and one requires some approximations. One such approximation is the large q limit which helps considerably in solving the Schwinger-Dyson equations for the SYK model being considered here \cite{Maldacena:2016hyu}. In order to consider this limit we first begin by noting that in the frequency space, the free fermion Green's function can be given as
\bal
G_0(\omega_n)=-\frac{1}{i \omega_n },~~ \omega_n =\frac{2\pi}{\beta}(n+1/2),\label{G0freew}
\eal
where $\omega_n$ is the Matsubara frequency. It is also possible to write the dispersionless free fermion Green's function in the $\tau$ space as
\be
G_0 (\t,\t') = \frac{1}{2} \sgn(\t-\t').\label{G0freet}
\ee 
such that in the large q limit one can recast the full Green's function $G(\t)$ as follows,
\be
G(\t,\t') = G_0 (\t,\t') \left[ 1 + \frac{1}{q} g(\t,\t') \right].\label{Gqexpn}
\ee

Given the large q expansion of Green's function above, one can in principle determine an effective action for the field $g(\t,\t')$ which is the $\cO(1/q)$ correction. For this, it may be noted that the saddle point self-energy in \eqref{sdGF} also scales with $1/q$ so that one can expand the determinant term in \eqref{IneffFin} as follows
\bal
\Tr \ln(\pa_{\t}-\S)=\Tr \ln(G_0^{-1})-\Tr\(G_0*\S\)-\frac{1}{2}\Tr\(G_0*\S*G_0*\S\)+\cdots,\label{detexpand}
\eal
where 
\bal
G_0^{-1}(\omega)\equiv \pa_{\t},\quad \Tr\(G_0*\S\)=\int d\t'' G_0(\t,\t'')\S(\t'',\t')
\eal

Substituting \eqref{Gqexpn} and \eqref{detexpand} in \eqref{IneffFin} and dropping the constant term $\Tr \ln(G_0^{-1})$ one can get the following form of the effective action in the large q limit
\bal
\frac{I^{(q)}_{eff}(J)}{N}&\simeq \frac{1}{4} \Tr\(G_0*\S*G_0*\S\)+\frac{1}{2q}\int\int d\t d\t'\S(\t,\t')G_0(\t,\t')g(\t,\t')\nn\\
&-\frac{\cJ^2}{4 q^2(1+\l)}\int \int d\t d\t'e^{g(\t,\t')}+\frac{\l}{N^{q-1}(1+\l)}\frac{\cJ^4}{16 q^4}\(\int \int d\t d\t'e^{g(\t,\t')}\)\(\int \int d\bar{\t}d\bar{\t}'e^{g(\bar{\t},\bar{\t}')}\).\label{geff}
\eal

Now on integrating out $\S(\t,\t')$ from the above action one can obtain an effective action for the $g(\t,\t')$ field. For this, we define the following field
\bal
\Phi(\t,\t')=\Tr\(G_0*\S\)=\int d\t'' G_0(\t,\t'')\S(\t'',\t'),\quad \pa_{\t}\Phi(\t,\t')=\S(\t,\t'),
\eal
which on substituting in \eqref{geff} gives us the following expression
\bal
\frac{I^{(q)}_{eff}(J)}{N}&\simeq \frac{1}{4} \Tr\(\Phi*\Phi\)+\frac{1}{2q}\int\int d\t d\t'\pa_{\t}\Phi(\t,\t')G_0(\t,\t')g(\t,\t')\nn\\
&-\frac{\cJ^2}{4 q^2(1+\l)}\int \int d\t d\t'e^{g(\t,\t')}\nn\\
&+\frac{\l}{N^{q-1}(1+\l)}\frac{ \cJ^4}{16 q^4}\(\int \int d\t d\t'e^{g(\t,\t')}\)\(\int \int d\bar{\t}d\bar{\t}'e^{g(\bar{\t},\bar{\t}')}\),\nn\\
&\simeq \frac{1}{4} \Tr\(\Phi*\Phi\)-\frac{1}{2q}\int\int d\t d\t'\Phi(\t,\t')\pa_{\t}\(G_0(\t,\t')g(\t,\t')\)\nn\\
&-\frac{\cJ^2}{4 q^2(1+\l)}\int \int d\t d\t'e^{g(\t,\t')}\nn\\
&+\frac{\l}{N^{q-1}(1+\l)}\frac{ \cJ^4}{16 q^4}\(\int \int d\t d\t'e^{g(\t,\t')}\)\(\int \int d\bar{\t}d\bar{\t}'e^{g(\bar{\t},\bar{\t}')}\),\label{geffmid}
\eal
where in writing the second line we have used integration by parts in the second term. Integrating over the field $\Phi(\t,\t')$ one can obtain the Liouville effective action for the field $g(\t,\t')$ as follows
\bal
\frac{I^{(q)}_{eff}(J)}{N}\simeq &\frac{1}{16q^2}\int\int d\t d\t'\pa_{\t}g(\t,\t')\pa_{\t'}g(\t,\t')-\frac{\cJ^2}{4 q^2(1+\l)}\int \int d\t d\t'e^{g(\t,\t')}\nn\\
&+\frac{\l}{N^{q-1}(1+\l)}\frac{ \cJ^4}{16 q^4}\(\int \int d\t d\t'e^{g(\t,\t')}\)\(\int \int d\bar{\t}d\bar{\t}'e^{g(\bar{\t},\bar{\t}')}\),\label{geffFin}
\eal
where in writing the above effective action from \eqref{geffmid} we have substituted the explicit value of $G_0(\t,\t')=1/2\sgn(\t-\t')$, such that it is only valid in the regime $\t>\t'$, for the integrals. From the effective action in \eqref{geffFin} one can obtain an equation of motion for the field $g(\t,\t')$ as
\bal
\pa_{\t}\pa_{\t'}g(\t,\t')=-\frac{2\cJ^2}{1+\l}\(1-\frac{\l}{2N^{q-1}} \frac{\cJ^2}{q^2}\int \int d\bar{\t}d\bar{\t}'e^{g(\bar{\t},\bar{\t}')}\) e^{g(\t,\t')},\label{geom}
\eal
where we will be interested in general solutions that are time translation invariant. This implies that the field $g(\t,\t')$ is a function of the difference of its arguments. 
\paragraph{Liouville field Theory:} Here we first define the following coordinates
\bal
\n=\t'-\t,\nn\\
\s=\t'+\t,\label{Lcoord}
\eal
which then helps us to write the effective action in \eqref{geffFin} as
\bal
\frac{I^{(q)}_{eff}(J)}{N}\simeq & \frac{1}{8q^2}\int\int d\n d\s\left(-\(\pa_{\n}g(\n,\s)\)^2+\(\pa_{\s}g(\n,\s)\)^2-\frac{4\cJ^2}{(1+\l)} e^{g(\n,\s)}\right.\nn\\
&\left.+\frac{\l}{N^{q-1}(1+\l)}\frac{ \cJ^4}{2 q^2}e^{g(\n,\s)}\(\int \int d\bar{\n}d\bar{\s}'e^{g(\bar{\n},\bar{\s})}\)\right),\label{geffFin1}
\eal
and the corresponding equations of motion as
\bal
-\pa^2_{\n}g(\n,\s)+\pa^2_{\s}g(\n,\s)=-\frac{2\cJ^2}{1+\l}\(1-\frac{\l}{2N^{q-1}} \frac{\cJ^2}{q^2}\int \int d\bar{\n}d\bar{\s}'e^{g(\bar{\n},\bar{\s})}\) e^{g(\n,\s)}.\label{geom1}
\eal

The above effective action for the non-Gaussian SYK model can now be considered as a two dimensional Liouville field theory. The dynamics or the quantization of this Liouville field theory depends only on the field $g(\n,\s)$ and it can be considered as the Weyl factor to the naive flat metric appearing in the kinetic term of the action \cite{Goel:2023svz}. Thus, one can motivate the following Euclidean two-dimensional anti-de-Sitter metric 
\bal
ds^2=e^{g(\n,\s)}(d\n^2-d\s^2),\label{adsMet}
\eal
or the Euclidean de-Sitter metric
\bal
ds^2=e^{g(\n,\s)}(-d\n^2+d\s^2),\label{dsMet}
\eal
for $g(\n,\s)$ satisfying the equations of motion in \eqref{geom1} \cite{Maldacena:2016hyu,Goel:2023svz}. As elucidated earlier, we are working with the Euclidean action \eqref{geffFin1} which does not indicate which metric to prefer. Thus, one may choose any one of the Euclidean de-Sitter or the anti-de-Sitter metrics described  above. This ambiguity is also due to the fact that in two dimensions de-Sitter and anti-de-Sitter spaces have the same isometries
and the same topology. However, here we first choose the anti-de-Sitter metric \eqref{adsMet} for which one can compute the Ricci scalar as
\bal
\cR=-e^{-g(\n,\s)}\(\pa^2_{\n}g(\n,\s)-\pa^2_{\s}g(\n,\s)\).\label{Ricci2ads}
\eal

Using the above form of the two-dimensional Ricci scalar one can recast the effective action in \eqref{geffFin1} in the following form
\bal
\frac{I^{(q)}_{eff}(J)}{N}\simeq &-\frac{1}{16q^2}\int\int d\n d\s\sqrt{-\tilde{g}}\left(\ln{\(-\tilde{g}(\n,\s)\)}\cR(\n,\s)+\frac{4\cJ^2}{(1+\l)} \right.\nn\\
&\left.-\frac{\l}{N^{q-1}(1+\l)}\frac{ \cJ^4}{ q^2}\(\int \int d\bar{\n}d\bar{\s}'\sqrt{-\tilde{g}}\)\right),\label{geffFin2ads}
\eal
where we have defined, $\tilde{g}=-e^{2g(\n,\s)}$ as the determinant of the corresponding AdS metric in \eqref{adsMet}. In a similar way when one considers the de-Sitter metric \eqref{dsMet} for which the Ricci scalar is given as
\bal
\cR=-e^{-g(\n,\s)}\(-\pa^2_{\n}g(\n,\s)+\pa^2_{\s}g(\n,\s)\),\label{Ricci2ds}
\eal
using which one can now recast the effective action in \eqref{geffFin1} as
\bal
\frac{I^{(q)}_{eff}(J)}{N}\simeq &\frac{1}{16q^2}\int\int d\n d\s\sqrt{-\tilde{g}}\left(\ln{\(-\tilde{g}(\n,\s)\)}\cR(\n,\s)-\frac{4\cJ^2}{(1+\l)} \right.\nn\\
&\left.+\frac{\l}{N^{q-1}(1+\l)}\frac{ \cJ^4}{q^2}\(\int \int d\bar{\n}d\bar{\s}'\sqrt{-\tilde{g}}\)\right),\label{geffFin2ds}
\eal

In the large q and N limit, the above effective action can then be used to compute the free energy and other thermodynamical properties  of the non-Gaussian SYK model which we address in the next subsection.

\subsection{Free Energy in the large q limit}  In order to determine the free energy, we begin by considering the partition function for the effective action in \eqref{geffFin1} as
\bal
\cZ(\cJ,\l)&=\int dg\;\exp{\(-\frac{I^{(q)}_{eff}(J)}{N}\)},\nn\\
&=\frac{1}{\sqrt{2\pi}}\int\limits_{-\infty}^{\infty} d\tilde{\cM}\;\exp{\(-\frac{\tilde{\cM}^2}{2}\)}\;\int dg\;\exp{\left[\frac{1}{8q^2}\(-\Theta_1+\(1+i\sqrt{\frac{2\l(1+\l)}{N^{q-1}}}\;\tilde{M}\)\frac{4\cJ^2}{\(1+\l\)}\Theta_0\)\right]},\label{VenZ}
\eal
where 
\bal
\Theta_0=\int\int d\n d\s\; e^{g\(\n,\s\)},\quad \Theta_1=\int\int d\n d\s\(-\(\pa_{\n}g(\n,\s)\)^2+\(\pa_{\s}g(\n,\s)\)^2\).\label{t1t0def}
\eal

One can further simplify the partition function by integrating out the $g$ field using a saddle point approximation as was done in \cite{Veneziano:1989hd}. For this, we first consider the equation of motion for the following action  
\bal
I_{eff}=-\frac{1}{8q^2}\(-\Theta_1+\(1+i\sqrt{\frac{2\l(1+\l)}{N^{q-1}}}\;\tilde{M}\)\frac{2\cJ^2}{\(1+\l\)}\Theta_0\),\label{It0t1}
\eal
which is given as
\bal
-\pa^2_{\n}g(\n,\s)+\pa^2_{\s}g(\n,\s)=-\(1+i\sqrt{\frac{2\l(1+\l)}{N^{q-1}}}\;\tilde{M}\)\frac{4\cJ^2}{\(1+\l\)}\;e^{g\(\n,\s\)},\label{REeq}
\eal

We are interested in finding a solution to the above equation of motion. For this, we consider the ansatz that the field $g(\n,\s)=g(\n)$ is a function of the $\n\in[0,\b]$ variable only. The integration over the $\s$ variable is also performed over the $[0,\b]$ interval. This gives us the following
\bal
I_{eff}=-\frac{\b}{8q^2} \int d\n\(\(\pa_{\n}g(\n)\)^2+\(1+i\sqrt{\frac{2\l(1+\l)}{N^{q-1}}}\;\tilde{M}\)\frac{4\cJ^2}{\(1+\l\)}e^{g(\n)}\),\label{It0t2}
\eal
where the equation of motion now takes the form
\bal
\pa^2_{\n}g(\n)=\(1+i\sqrt{\frac{2\l(1+\l)}{N^{q-1}}}\;\tilde{M}\)\frac{2\cJ^2}{\(1+\l\)}\;e^{g\(\n\)},\label{REeq2}
\eal

For the boundary conditions $g(0)=g(\b)=0$, The solution to the above equation of motion can be given as 
\bal
g(\n)=2\log{\left[\frac{\cos\left(\frac{\pi  \a }{2}\right)}{\cos\left(\pi  \a  \left(\frac{1}{2}-\frac{|\n|}{\beta }\right)\right)}\right]},\quad \pi^2 \a^2=\(1+i\sqrt{\frac{2\l(1+\l)}{N^{q-1}}}\;\tilde{M}\)\frac{\cJ^2\b^2}{\(1+\l\)} \cos ^2\left(\frac{\pi  \a }{2}\right).\label{gsoln}
\eal
where the parameter $\a$ ranges from zero to one as $\b\cJ$ ranges from zero to infinity. The parameter $\cJ$ is dimensionful and remains fixed as q tends towards larger values.

Now in order to determine the on-shell value of the effective action $I_{eff}$, we consider its derivative with respect to $\cJ$ while keeping the parameters $q,\b,\l$ and $\tilde{M}$ fixed as
\bal
\cJ\pa_{\cJ}I_{eff}&=\frac{\a}{1+\frac{\pi\a}{2}\tan{\(\frac{\pi\a}{2}\)}}\pa_{\a}I_{eff},\nn\\
&=-\frac{\b}{2q^2} \(1+i\sqrt{\frac{2\l(1+\l)}{N^{q-1}}}\;\tilde{M}\)\frac{\cJ^2}{\(1+\l\)}\int d\n\;e^{g(\n)}=-\frac{\b}{4q^2}(g'(\b)-g'(0)),\nn\\
&=-\frac{\pi  \a }{q^2} \tan \(\frac{\pi  \a }{2}\),\label{It0t3}\\
\pa_{\a}I_{eff}&=-\frac{\pi}{q^2} \tan \(\frac{\pi  \a }{2}\)\(1+\frac{\pi\a}{2}\tan{\(\frac{\pi\a}{2}\)}\),\label{It0t4}
\eal
where in writing the above expressions it may be noted that we have used the second relation in \eqref{gsoln} to turn the $\cJ$ derivative into $\a$ derivative. One can now determine $I_{eff}$ by simply integrating \eqref{It0t4} with respect $\a$ which gives us
\bal
I_{eff}=-\frac{1}{2}\log{2}-\frac{\pi\a}{q^2}\(\tan{\(\frac{\pi\a}{2}\)}-\frac{\pi\a}{4}\),\label{freeI}
\eal
where the integration constant can be fixed using the fact that for $\cJ\to 0$, it is simply the log of the total dimension of the Hilbert space.

Using the expression for $I_{eff}$ in \eqref{VenZ} renders the following form for the partition function in the finite temperature regime 
\bal
\cZ(\cJ,\l)=\frac{1}{\sqrt{2\pi}}\int\limits_{-\infty}^{\infty} d\tilde{\cM}\;e^{-\cI(\tilde{M})},\label{VenZ0}
\eal
 where
\bal
\cI(\tilde{M})&=\frac{\tilde{\cM}^2}{2}-\frac{1}{2}\log{2}-\frac{\pi\;\a\(\tilde{M}\)}{q^2}\(\tan{\(\frac{\pi\;\a\(\tilde{M}\)}{2}\)}-\frac{\pi\;\a\(\tilde{M}\)}{4}\),\label{VenzIM}
\eal
and $\a$ is considered to be a function of the auxiliary variable $\tilde{M}$ using the constraint relation in \eqref{gsoln}. In order to determine the saddle point value for the auxiliary variable $\tilde{M}$ we consider the zeros of the following 
\bal
\cI'(\tilde{M})=\tilde{M}+\frac{i \pi  \alpha  \tan \left(\frac{\pi  \alpha }{2}\right) \sqrt{\frac{2\l(1+\l)}{N^{q-1}}}}{2q^2 \left(1+ i\tilde{M} \sqrt{\frac{2\l(1+\l)}{N^{q-1}}}\right)},\label{IpM}
\eal
which is trivial to obtain by determining $\a'(\tilde{M})$ from the constraint relation in \eqref{gsoln} and then using the same to replace $\cJ$ in terms of $\a$ and $\tilde{M}$. Setting $\cI'(\tilde{M})$ to zero gives the following value for the auxiliary field 
\bal
\tilde{M}_0=\frac{i}{\sqrt{\frac{8\l(1+\l)}{N^{q-1}}}}\(1\pm \sqrt{1+\frac{  4 \pi \a \l  (1+\l) \tan \left(\frac{\pi  \alpha }{2}\right)}{q^2 N^{q-1}}}\).\label{Msad}
\eal
which when combined with the constraint in \eqref{gsoln} gives us
\bal
\b\cj=\frac{\pi  \a  \sqrt{2 (1+\l)}}{ \cos \left(\frac{\pi  \a }{2}\right)\sqrt{1\mp\sqrt{1+\frac{4 \pi  \a \l (1+\l)   \tan \left(\frac{\pi  \a }{2}\right)}{q^2 N^{q-1}}}}}.\label{gconstM}
\eal
such that for the solution of $\tilde{M}$ with the positive sign we get the corresponding negative sign in the above expression.

The partition function \eqref{VenZ0} can now be evaluated using a saddle point approximation around the solution \eqref{Msad} with the constraint \eqref{gconstM} as 
\bal
\cZ(\cJ,\l)\simeq &\frac{1}{\sqrt{\cI''(\tilde{M}_0)}} e^{-\cI(\tilde{M}_0)},\label{VenZ1}
\eal
where
\bal
\cI(\tilde{M}_0)&=-\frac{1}{2}\log{2}+\frac{\pi ^2 \a^2}{4 q^2}-\frac{\pi  \a}{q^2}\tan \(\frac{\pi  \a }{2}\)\nn\\
&+\frac{N^{q-1}}{16 \l  (1+\l)}\(1-\sqrt{1+\frac{  4 \pi \a \l  (1+\l) \tan \left(\frac{\pi  \alpha }{2}\right)}{q^2 N^{q-1}}}\)^2,\\
\cI''(\tilde{M}_0)&=1+\frac{ \pi  \a \l (1+\l)\sec ^2\left(\frac{\pi  \a }{2}\right) (\pi  \a  \cos (\pi  \a )-\sin (\pi  \a ))}{N^{q-1}\left(1+\frac{\pi  \a}{2}  \tan \left(\frac{\pi  \a }{2}\right)\right) \left(1+\sqrt{1+\frac{4 \pi  \a \l (1+\lambda) \tan \left(\frac{\pi  \alpha }{2}\right)}{q^2 N^{q-1}}}\right)^2}.\label{ffunc}
\eal
which we have evaluated for the constraint in \eqref{Msad} for the negative sign. The free energy can now be determined from \eqref{VenZ1} as
\bal
-\b F=\log{\cZ}&=-\frac{\pi ^2 \a^2}{4 q^2}+\frac{\pi  \a}{ q^2}\tan \(\frac{\pi  \a }{2}\)-\frac{N^{q-1}}{16 \l  (1+\l)}\(1-\sqrt{1+\frac{  4 \pi \a \l  (1+\l) \tan \left(\frac{\pi  \alpha }{2}\right)}{q^2 N^{q-1}}}\)^2\nn\\
&-\frac{1}{2}\log\left[\frac{1}{2}+\frac{ \pi  \a \l (1+\l)\sec ^2\left(\frac{\pi  \a }{2}\right) (\pi  \a  \cos (\pi  \a )-\sin (\pi  \a ))}{2N^{q-1}\left(1+\frac{\pi  \a}{2}  \tan \left(\frac{\pi  \a }{2}\right)\right) \left(1+\sqrt{1+\frac{4 \pi  \a \l (1+\lambda) \tan \left(\frac{\pi  \alpha }{2}\right)}{q^2 N^{q-1}}}\right)^2}\right],\label{FreeVen}
\eal
for the modified constraint in \eqref{gconstM} with the positive sign. Moreover, in the large $N$ limit we can expand the free energy as
\bal
-\b F=&\frac{1}{2}\log{2}+\frac{\pi\a}{q^2}\(\tan\(\frac{\pi\a}{2}\)-\frac{\pi\a}{4}\)+\frac{\l (1+\l)}{N^{q-1}}\left(\frac{\pi ^2 \a ^2 }{4 q^4} \tan ^2\left(\frac{\pi  \a }{2}\right)\right.\nn\\
&\left.+\frac{\pi\a}{2 q^2}\frac{\pi  \alpha  \cos (\pi  \alpha )-\sin (\pi  \alpha )}{\cos ^2\left(\frac{\pi  \alpha }{2}\right) \left(2+\pi  \alpha  \tan \left(\frac{\pi  \alpha }{2}\right)\right)}\right)+\cO(1/N^{2q-2})
\eal
 where the first two terms yield the free energy for the SYK model with Gaussian disorder. The other terms in the large N expansion of the free energy represent the $\l$ dependent order $\cO(1/N^{2q-2})$ corrections signaling deviations away from Gaussianity.

\subsection{Coleman Mechanism for a small cosmological constant}
The effective action obtained in \eqref{geffFin2ds} for the case of Euclidean de-Sitter metric is surprisingly reminiscent of the non-local toy model considered by  Veneziano in \cite{Veneziano:1989hd}. This model was used to provide a new mechanism for suppressing the cosmological constant which is related to Coleman's \cite{Coleman:1988tj} idea of considering the effects of the wormhole solutions in Euclidean quantum gravity. It is well known in the literature that a two-site coupled SYK model possesses a wormhole solution that violates the null energy condition \cite{Maldacena:2018lmt}. Thus one may propose a Coleman like mechanism for two-site coupled SYK model whose low energy physics is captured by the bimetric JT gravity\cite{Maldacena:2018lmt}. However, in the present case, we observed that on considering non-Gaussian disorder in the SYK model one obtains a non-local action as in \eqref{geffFin1}. This effective action may then be used to formulate a mechanism for suppressing the cosmological constant. This suppression may work here since as shown in \cite{Polchinski:1988ua}, $d=2\, (\text{mod}\, 4)$ is required which is the case here.

To elucidate the given mechanism for the case of the non-Gaussian SYK model, we begin by considering the effective action in \eqref{geffFin1} which after performing integration over $\s\in[0,\b]$ coordinate for the ansatz $g(\n,\s)=g(\n)$, can be written as
\bal
\frac{I^{(q)}_{eff}(J)}{N}\simeq &\frac{\b}{8q^2}\int d\n \left(-\(\pa_{\n}g(\n)\)^2-\frac{4\cJ^2}{(1+\l)} e^{g(\n)}-\frac{\l\b}{N^{q-1}(1+\l)}\frac{ \cJ^4}{2 q^2}e^{g(\n)}\int d\bar{\n}e^{g(\bar{\n})}\right),\nn\\
=&-\b\L_0\theta_0+\b\L_1\theta_1+\frac{\b^2 C}{2}\theta_0^2,\label{geffVenZ}
\eal
where in writing the second line in the above expression, we have identified the following 
\bal
\theta_0&=\int d\n e^{g(\n)},\quad \theta_1=-\int d\n \(\pa_{\n}g(\n)\)^2,\nn\\
\L_0&=\frac{\cJ^2}{2q^2(1+\l)},\quad \L_1=\frac{1}{8q^2},\quad C=\frac{\l}{N^{q-1}(1+\l)}\frac{ \cJ^4}{8 q^4},\label{L0L1C}
\eal
and have considered that the field, $g(\n,\s)=g(\n)$ is a function of the $\n$ variable only. The constants $\L_0$ and $\L_1$ are positive definite whereas $C$ is a very small positive constant. For the de-Sitter geometry, using integration by parts the parameter $\theta_1$ is related to the Ricci scalar in \eqref{Ricci2ds} in the following way
\bal
\theta_1=\int d\n \;g(\n)\pa^2_{\n}g(\n)=\int d\n \sqrt{-\tilde{g}} \ln{\(-\tilde{g}(\n)\)}\cR(\n)
\eal
where $\tilde{g}=-e^{2g(\n)}$, is the determinant of the corresponding Euclidean de-Sitter metric. The nonlocal effective action in \eqref{geffVenZ} is very similar to the one studied in \cite{Veneziano:1989hd} with the only difference being that in our case we are restricted to two dimensions only. For this effective action, one can write down the following partition function
\bal
\cZ(\cJ,\l)&=\int dg\;\exp{\(-\frac{I^{(q)}_{eff}(J)}{N}\)},\nn\\
&=\frac{1}{\sqrt{2\pi C}}\int\limits_{-\infty}^{\infty} d\tilde{\cM}\;\exp{\(-\frac{\tilde{\cM}^2}{2 C}\)}\;\int dg\;\exp{\left[\b\((\L_0+i \tilde{M})\theta_0-\L_1\theta_1\)\right]},\label{VenZM}
\eal
which is a bit different from the partition function considered in \eqref{VenZ} where the integration contour over the auxiliary variable $\tilde{M}$ runs along the real axis instead of the imaginary. The above partition function can be evaluated by integrating out the $g$ field using a saddle point approximation for which we first consider the effective action
\bal
I_{eff}(\tilde{M})=-\b(\L_0+i \tilde{M})\theta_0+\b\L_1\theta_1
\eal
whose corresponding equation of motion is further given by
\bal
\pa^2_{\n}g(\n)=\(\frac{\L_0+i \tilde{M}}{2\L_1}\)e^{g(\n)}
\eal

For the boundary conditions $g(0)=g(\b)=0$, once again the solution to the above equation of motion can be given as 
\bal
g(\n)=2\log{\left[\frac{\cos\left(\frac{\pi  \a }{2}\right)}{\cos\left(\pi  \a  \left(\frac{1}{2}-\frac{|\n|}{\beta }\right)\right)}\right]},\label{gsolMn}
\eal
for the constraint
\bal
\frac{\pi \a}{\sec\left(\frac{\pi  \a }{2}\right)}=\b\sqrt{\frac{\L_0+i \tilde{M}}{4\L_1}},\quad \L_{eff}=\L_0+i \tilde{M}\label{constgsolMn}
\eal
where the parameter $\a$ ranges from zero to one as the quantity under square root ranges from zero to infinity. Given the solution for $g(\n)$ above, one can determine the on-shell value of the effective action $\tilde{I}(\tilde{M})$ by considering its derivative with respect to $\L_{eff}$ while keeping the parameters $\L_1$ and $\tilde{M}$ fixed as
\bal
2\L_{eff}\pa_{\L_{eff}}I_{eff}(\tilde{M})&=\frac{\a}{1+\frac{\pi\a}{2}\tan{\(\frac{\pi\a}{2}\)}}\pa_{\a}\tilde{I}(\tilde{M}),\nn\\
&=-2\b\(\L_0+i \tilde{M}\)\int d\n\;e^{g(\n)}=-4\b\L_1(g'(\b)-g'(0)),\nn\\
&=-16\pi\L_1 \a  \tan \(\frac{\pi  \a }{2}\),\label{IML1a}\\
\pa_{\a}I_{eff}(\tilde{M})&=-16\pi \L_1\tan \(\frac{\pi  \a }{2}\)\(1+\frac{\pi\a}{2}\tan{\(\frac{\pi\a}{2}\)}\),\label{IMa}
\eal
where in writing the above expressions it may be noted that we have used the relation in \eqref{constgsolMn} to turn the $\L_0$ derivative into $\a$ derivative. Now the on-shell value of $I_{eff}(\tilde{M})$ can be determined by simply integrating \eqref{IMa} with respect to $\a$ which gives us
\bal
I_{eff}(\tilde{M})=-\frac{1}{2}\log{2}-16\pi\L_1\a\(\tan{\(\frac{\pi\a}{2}\)}-\frac{\pi\a}{4}\),\label{freeIV}
\eal
which is different from the on-shell action obtained previously in \eqref{freeI} with different parametrizations. We now explore the low temperature $(\b\to \infty)$ limit of the given effective action. In this limit where $\a\to 1$, one can use the constraint in \eqref{constgsolMn} to obtain the following
\bal
\a=1-\frac{2}{\b\sqrt{\frac{\L_0+i \tilde{M}}{4\L_1}}}+\frac{4}{\b^2\(\frac{\L_0+i \tilde{M}}{4\L_1}\)}-\frac{(24+\pi ^2)}{3 \b^3\(\frac{\L_0+i\tilde{M}}{4\L_1}\)^{3/2}}+\cdots,\label{aconstL}
\eal

The effective action in \eqref{freeIV} is valid for any non-zero finite temperature. However, using \eqref{aconstL} one can determine the expression for the effective action in the low temperature regime as
\bal
I_{eff}(\tilde{M})=-\(\frac{1}{2}\log{2}-4 \pi ^2 \L_1\)-8 \b \sqrt{ \L_1\(\L_0+i \tilde{M}\)}-\frac{16 \pi ^2 \L_1^{3/2}}{\beta \sqrt{\L_0+i\tilde{M}}},\label{lowTFreeIV}
\eal
which gives us the partition function in \eqref{VenZM} as
\bal
\cZ(\cJ,\l)=\frac{1}{\sqrt{2\pi C}}\int\limits_{-\infty}^{\infty} d\tilde{\cM}\;e^{-\cI(\tilde{M})},\label{VenZMLT}
\eal
where
\bal
\cI(\tilde{M})=\frac{\tilde{M}^2}{2 C}-\(\frac{1}{2}\log{2}-4 \pi ^2 \L_1\)-8 \b \sqrt{ \L_1\(\L_0+i\tilde{M}\)}-\frac{16 \pi ^2 \L_1^{3/2}}{\beta \sqrt{\L_0+i\tilde{M}}}.\label{IMVenZ}
\eal

The saddle point value for the auxiliary variable $\tilde{M}$ can be determined from the zeros of the following
\bal
\cI'(\tilde{M})=\frac{\tilde{M}}{C}-i \sqrt{ \frac{\L_1}{\L_0+i\tilde{M}}}\(4 \b -\frac{8 \pi ^2}{\b}\(\frac{\L_1}{ \L_0+i\tilde{M}}\)\),\label{lowTFreeIVp}
\eal
where it is not clear which value of $\tilde{M}$ obtained from equating the above to zero gives the physical solutions. For this, we first consider the case when $\tilde{M}$ is small which gives us
\bal
\cI'(\tilde{M})\simeq \frac{\tilde{M}}{C}-i\sqrt{ \frac{\L_1}{\L_0}}\(4 \b-\frac{8 \pi ^2\L_1}{\b\L_0}\),\label{SMappxVIp}
\eal
and on imposing $\cI'(\tilde{M})=0$, one obtains the following saddle point value of the auxiliary variable
\bal
\tilde{M}_{0}=i4\b C\sqrt{ \frac{\L_1}{\L_0}}\(1-\frac{2 \pi ^2\L_1}{\b^2\L_0}\)\simeq i \frac{\l\b\cJ^3}{4 q^4 N^{q-1}\sqrt{\lambda +1} },\label{SsadM}
\eal

This saddle point corresponds to small $\tilde{M}$ and a small volume universe which can be approximated as
\bal
\cV=\int\limits_{0}^{\b}d\n e^{g(\n)}\simeq 4\sqrt{ \frac{\L_1}{\L_0}}=2\frac{\sqrt{1+\l}}{\cJ}.\label{SsadV}
\eal
and in a similar way, one can also approximate $\cI(\tilde{M})$ as
\bal
\cI(\tilde{M}_0)\simeq -\frac{\log (2)}{2}+\frac{\pi ^2}{2 q^2}-\frac{2 \beta  \cJ}{q^2\sqrt{1+\lambda} }-\frac{\pi ^2 \sqrt{1+\lambda}}{\beta  \cJ q^2}+\frac{\l \left(\pi ^2 (1+\l)-2 \b^2 \cJ^2\right)^2}{16 \b^2 \cJ^4 q^4 N^{q-1}}+\cO\(\frac{1}{N^{2q-2}}\)
\eal
which suggests that as the volume $\cV\propto 1/\cJ\to \infty$ the action $\cI(\tilde{M}_0)$ also diverges implying that this saddle point does not correspond to a stable two-dimensional universe.

The second saddle point corresponds to the small effective cosmological constant $\L_{eff}=\L_0+\tilde{M}$ and a large volume universe,
\bal
\L_{eff}=4 \L_1 \left(\frac{\pi ^2 C}{\b \L_0}\right)^{2/3}=\frac{1}{4}\(\frac{\pi^4\l^2\cJ^4}{2 N^{2q-2}q^{10}\b^2}\)^{1/3}\ll \L_0,\quad \cV\simeq 2\left(\frac{\b \L_0}{ \pi^2 C}\right)^{1/3}+\frac{\left(\pi ^2-24\right) \Lambda _0}{3\pi^2 \b C},\label{SLEM&V}
\eal
Notice that $\beta$ is related to the size of the Universe, as it is coming from the integration over the $\sigma$ coordinate. As a result, the suppression of the cosmological constant is coming from several factors namely, the large volume as $\beta \rightarrow \infty$, the large $N \gg1$ limit and large values of the parameter $q$ which has positive exponent in the expression of $\L_{eff}$ in \eqref{SLEM&V}. We also compute the free energy as
\bal
-\b F=\log{\cZ}=\frac{1}{2} \left(\frac{\L_0^2}{C}-8 \L_1 \left(\frac{\pi ^2}{\b}-\frac{\pi^{4/3} \sqrt[3]{\L_0}}{\beta ^{4/3} \sqrt[3]{C}}\right)-\log \left(\frac{3 \beta ^{4/3} \L_0^{5/3}}{8 \pi ^{4/3} C^{5/3} \L_1}\right)\right)
\eal
which on using the relations in \eqref{L0L1C} can be shown to remain finite as $\b\to \infty$, implying that the second saddle is the physical one. This suppression of the cosmological constant for the non-local gravitational effective action arising from the large-$q$ description of the non-Gaussian SYK model, is in stark parallel with the case of non-local gravitational action considered in \cite{Veneziano:1989hd} to demonstrate the Coleman-Veneziano mechanism for a small cosmological constant.

\section{Schwarzian theory with non-Gaussian disorder}
 In the previous sections, we studied the SYK model in the large N limit and obtained a bilocal effective action in terms of the fields $\Sigma$ and $G$ in \eqref{IneffFin} using a non-Gaussian disorder averaging. Here we wish to look specifically at the properties of the non-Gaussian SYK model in the regime $1\ll\b\cJ\ll N$, that is in the large N and low-temperature limit. However, before we proceed we would like to express the partition function corresponding to this bilocal effective action through a Hubbard-Stratonovich transformation as
 \bal
\cZ(\l,\cJ)=\frac{1}{\sqrt{2\pi}}\int\limits_{-\infty}^{\infty} d\tilde{M} \exp{\[-\frac{\cI(\S,G,\tilde{M})}{N}\]}
 \eal
where
\bal
-\frac{\cI(\S,G,\tilde{M})}{N}&=\frac{1}{2} \Tr\ln \(\d(\t-\t')\pa_{\t}-\S(\t,\t')\)-\frac{1}{2}\int\int d\t d\t'\S(\t,\t')G(\t,\t')\nn\\
&-\frac{\tilde{M}^2}{2}+\(1+i\sqrt{\frac{2\l(1+\l)}{N^{q-1}}}\;\tilde{M}\)\frac{2^{q-2}\cJ^2}{ q^2\(1+\l\)}\int \int d\t d\t'G(\t,\t')^{q},\label{IMeff}
\eal
such that on integrating over the auxiliary field $\tilde{M}$ one recovers precisely the effective action mentioned in \eqref{IneffFin} before. For the effective action \eqref{IMeff} the saddle point equations of motion for the bilocal fields $G(\t,\t')$ and $\S(\t,\t')$ can now be given as 
\bal
&G(\t,\t')=\(\d(\t-\t')\pa_{\t}-\S(\t,\t')\)^{-1},\label{sdSFM}\\
&\S(\t,\t')=\(1+i\sqrt{\frac{2\l(1+\l)}{N^{q-1}}}\;\tilde{M}\)\frac{2^{q-1}\cJ^2}{ q(1+\l)}G(\t,\t')^{q-1},\label{sdGFM}
\eal

In the low-temperature regime, it is to be noted that the kinetic term acts as an irrelevant perturbation to the action in \eqref{IMeff} and thus one can drop it from the action and equations of motion in \eqref{sdGFM} and \eqref{sdSFM} to obtain the following
 \bal
&\int d\t'' G(\t,\t'')\;\S(\t'',\t')=-\d(\t-\t').\label{sdSFMLT}\\
&\S(\t,\t')=\(1+i\sqrt{\frac{2\l(1+\l)}{N^{q-1}}}\;\tilde{M}\)\frac{2^{q-1}\cJ^2}{ q(1+\l)}G(\t,\t')^{q-1},\label{sdGFMLT}
 \eal

Interestingly, dropping the kinetic term from the action and the equations of motion has an important consequence in the sense that they both now have an emergent reparameterization symmetry i.e., to say they are invariant under the following transformations
\bal
G(\t,\t')&\to \[\phi'(\t)\phi'(\t')\]^{\D} G\(\phi(\t),\phi(\t')\),\nn\\
\S(\t,\t')&\to \[\phi'(\t)\phi'(\t')\]^{\D(q-1)} \S\(\phi(\t),\phi(\t')\),\label{transGS}
\eal
 where $\D=1/q$, is the scaling dimension of the Fermionic field $\psi_{i}(\t)$. In the large N limit, one can solve the equations of motion in \eqref{sdSFM} and \eqref{sdGFM} numerically. However, due to the reparameterization symmetry in the low-temperature limit, one can obtain a solution for $G$ that satisfies the equations of motion \eqref{sdSFMLT} and \eqref{sdGFMLT}. It has the form of a conformal two-point function which can be given as
 \bal
 G_{c}(\t,\t')=\frac{b}{\left|\t-\t'\right|^{2\D}}\sgn{(\t-\t')},\quad  G^{\phi}_{c}(\t,\t')=b\(\frac{\sqrt{\phi'(\t)\phi'(\t')}}{\abs{\phi(\t)-\phi(\t')}}\)^{2\D}\label{conG}
 \eal
where the second solution above corresponds to the transformation $t\to\phi(\t)$ and b is a constant. One can also obtain a finite temperature solution using $\phi(\t)=\tan\left(\frac{\pi\t}{\b}\right)$ in \eqref{transGS} as
 \bal
 G_{c}(\t,\t')=b\[\frac{\pi}{\b\sin\(\frac{\pi(\t-\t')}{\b}\)}\]^{2\D}\sgn{(\t-\t')},\label{conGb}
 \eal

Furthermore, using the conformal two-point function given above one can also obtain the self-energy in \eqref{sdGFMLT}  as
 \bal
  \S_{c}(\t,\t')&=\(1+i\sqrt{\frac{2\l(1+\l)}{N^{q-1}}}\;\tilde{M}\)\frac{2^{q-1}\cJ^2}{ q(1+\l)}\;G_c(\t,\t')^{q-1},\label{conS}
 \eal
 where the subscript $c$ implies that it is evaluated at the conformal point in the low-temperature regime. For the conformal two-point function the unknown constant $b$ is determined by using \eqref{conG} or \eqref{conGb} in the expression for the self-energy \eqref{sdSFMLT}, which subsequently gives us
 \bal
\(1+i\sqrt{\frac{2\l(1+\l)}{N^{q-1}}}\;\tilde{M}\)\frac{2^{q-1}\cJ^2}{ q(1+\l)} b^{q}\pi=\(\frac{1}{2}-\frac{1}{q}\)\tan\(\frac{\pi}{q}\),\quad \D=\frac{1}{q}.\label{conGpara}
 \eal

In the previous section the effective action \eqref{IneffFin} for the SYK model with non-Gaussian disorder averaging was obtained by using the steepest descent method through the replica diagonal saddle. Next, we expressed it as \eqref{IMeff} using a Hubbard-Stratonovich transformation. Then in the low-temperature regime $1\ll\b\cJ\ll N$, we showed that the action \eqref{IMeff} and the equations of motion \eqref{sdGFMLT}-\eqref{sdSFMLT} without the kinetic term are invariant under a time reparametrization, $\t\to \phi(\t)$. This fact permitted us to obtain the conformal saddle point solutions for the fields $G$ and $\S$ as given by \eqref{conGb} and \eqref{conS} respectively. However, it may be observed that the conformal saddle point solutions remain invariant under M$\ddot{o}$bius or $SL(2,R)$ transformations indicating the spontaneous breaking of the conformal/reparameterization symmetry down to global $SL(2,R)$ subgroup. On the other hand, if we write the equation of motion in \eqref{sdSFMLT} as
\bal
&\int d\t'' G(\t,\t'')\;\S(\t'',\t')+\int d\t''G(\t,\t'')s(\t'',\t')=-\d(\t-\t'),\quad s(\t,\t')=-\pa_{\t}\d(\t-\t'),\label{sdSFgen}
\eal
then they can be used to write the on-shell effective action ($\cS=\cI^{on-shell}$) from \eqref{IMeff} in the following way
\bal 
\frac{\cS}{N}=&\frac{1}{2} \Tr\ln \(G\)+\frac{1}{2}\int d\t\left.\pa_{\t}G(\t,\t')\right|_{\t'=\t}+\frac{\tilde{M}^2}{2}\nn\\
&-\(1+i\sqrt{\frac{2\l(1+\l)}{N^{q-1}}}\;\tilde{M}\)\frac{2^{q-2}\cJ^2}{ q^2(1+\l)}\int \int d\t d\t'G(\t,\t')^{q},\nn\\
=&\frac{1}{2} \Tr\ln \(G\)-\frac{1}{2}\int\int d\t d\t'G(\t,\t')s(\t,\t')+\frac{\tilde{M}^2}{2}\nn\\
&-\(1+i\sqrt{\frac{2\l(1+\l)}{N^{q-1}}}\;\tilde{M}\)\frac{2^{q-2}\cJ^2}{ q^2(1+\l)}\int \int d\t d\t'G(\t,\t')^{q},\label{Seff}
\eal
as argued in \cite{Das:2020kmt,Jevicki:2016bwu,Jevicki:2016ito}. It may be observed that the source term including $s(\t,\t')$, explicitly breaks the conformal invariance. It is this source term that governs the dynamics of the reparameterization modes $\phi(\t)$ as it lifts the degeneracy of the solutions for the action \eqref{IMeff} without the kinetic term and ensures that there is a single saddle. We now wish to consider the effective action for the fluctuations around the conformal solutions in \eqref{conG} and \eqref{conGb} given before. For this one can expand the field $G$ around the conformal solution as follows
\bal
G(\t,\t')=G_{c}(\t,\t')+g(\t,\t').\label{GSfluc}
\eal

Using the above expansion, the action in \eqref{Seff} can be written to the second order in $g$ as,
\bal
\frac{\cS}{N}&=-\frac{1}{2}\int\int d\t_1 d\t_2 \;G_c(\t_1,\t_2)s(\t_1,\t_2)-\frac{1}{2}\int\int d\t_1 d\t_2\;g(\t_1,\t_2)s(\t_1,\t_2)+\cS_{c}[G_c]\nn\\
&+\frac{1}{4}\int d\t_1 d\t_2d\t_3d\t_4 \; g(\t_1,\t_2)\; \cK(\t_1,\t_2;\t_3,\t_4) \;g(\t_3,\t_4),\label{geff1}
\eal
where
\bal
\cS_{c}[G]=&\frac{1}{2} \Tr\ln \(G\)+\frac{\tilde{M}^2}{2}-\(1+i\sqrt{\frac{2\l(1+\l)}{N^{q-1}}}\;\tilde{M}\)\frac{2^{q-2}\cJ^2}{ q^2(1+\l)}\int \int d\t d\t'G(\t,\t')^{q},\label{Gceff}
\eal
and the kernel $\cK(\t_1 , \t_2 ; \t_3 , \t_4 )$  is simply defined as
\bal
\cK(\t_1 , \t_2 ; \t_3 , \t_4 )=&\frac{\d^2\cS_c(G_c)}{\d G_c(\t_1 , \t_2)\;\d G_c( \t_3 , \t_4)},\nn\\
=& G_c^{-1}(\t_1 , \t_3)G_c^{-1}(\t_2 , \t_4)\nn\\
&+\(1+i\sqrt{\frac{2\l(1+\l)}{N^{q-1}}}\;\tilde{M}\)\frac{2^{q-1}(q-1)\cJ^2}{q(1+\l)}G_c^{q-2}(\t_1 , \t_2)\d(\t_1,\t_3)\d(\t_2,\t_4),\label{kerGc}
\eal
where $G_{c}^{-1}(\t,\t')=-\S_{c}(\t,\t')$, which is trivial to observe from the equations of motion in \eqref{sdSFMLT} in the low-temperature approximation.

Varying the action \eqref{geff1} with respect to $g(\t,\t')$ shows that it satisfies the following equation of motion
\bal
\int d\t_3d\t_4\; \ck(\t_1 , \t_2 ; \t_3 , \t_4 )\;g(\t_3,\t_4)=s(\t_1,\t_2),\label{kerg}
\eal
and one can use this as an identity to evaluate the first term in \eqref{geff} which breaks the reparametrization symmetry of the action explicitly. From the form of the kernel above it may be seen that it transforms under a scaling as $\cK\sim |\t|^{-4+4/q} $. This may prompt one to consider that $g(\t,\t')$  has the form $g(\t_1,\t_2)\propto \sgn{(\t_{12})|\t_{12}|^{-\frac{4}{q}}}$, 
 which contradicts with the desired $\d'$-source of \eqref{kerg}. This clearly implies that the $\d'$-source can only be matched at the non-perturbative level as argued in \cite{Das:2020kmt, Jevicki:2016bwu}. In order to rectify this one can consider an ansatz for the first order perturbative field $g(\t,\t')$ as
\bal
g(\t_1,\t_2)=\cB \frac{\sgn{(\t_{12})}}{|\t_{12}|^{\frac{2}{q}+2h}}
\eal
where $h$ is a positive definite parameter and $\cB$ is a normalization constant that is fixed numerically as demonstrated in \cite{Jevicki:2016bwu}. With such an ansatz for the $g(\t_1,\t_2)$ field one can now evaluate the left hand side of the \eqref{kerg}. For the first term in the expression of $\ck$, the integral on the left hand side in \eqref{kerg} is given as
\bal
&\int d\t_3d\t_4  G_c^{-1}(\t_1 , \t_3)G_c^{-1}(\t_2 , \t_4) g(\t_3,\t_4)\nn\\
&=\[\(1+i\sqrt{\frac{2\l(1+\l)}{N^{q-1}}}\;\tilde{M}\)\frac{2^{q-1}\cJ^2}{ q(1+\l)}\]^2b^{2q-2}\int d\t_3 d\t_4  \frac{{\rm sgn}(\t_{13}) \, {\rm sgn}(\t_{24}) \, {\rm sgn}(\t_{34})}{|\t_{13}|^{2\(1-\frac{1}{q}\)} \, |\t_{24}|^{2\(1-\frac{1}{q}\)} \, |\t_{34}|^{2\(h+\frac{1}{q}\)}}
\eal
which can be further evaluated by using the following integral 
\begin{align}
		\int d\t_3 d\t_4 \ \frac{{\rm sgn}(\t_{13}) \, {\rm sgn}(\t_{24}) \, {\rm sgn}(\t_{34})}{|\t_{13}|^{2\Delta} \, |\t_{24}|^{2\Delta} \, |\t_{34}|^{2\alpha}}
		=&\, - \pi^2 \left[ \frac{\sin(2\pi \alpha)\, + \, 2\sin(2\pi (\alpha+\Delta))
		\, + \, \sin(2\pi (\alpha+2\Delta))}{\sin(2\pi \alpha)\sin(2\pi \Delta)\sin(2\pi (\alpha+\Delta))\sin(2\pi (\alpha+2\Delta))} \right] \nonumber\\
		&\qquad \times \frac{\Big[ \sin(2\pi \Delta) + \sin(2\pi (\alpha+\Delta)) \Big] \Gamma(1-2\Delta)}{\Gamma(2\alpha)\Gamma(2\Delta)\Gamma(3-2\alpha-4\Delta)}
		\frac{{\rm sgn}(\t_{12})}{|\t_{12}|^{2\alpha+4\Delta-2}} \, .
	\label{Polchinski-integrals}
	\end{align}
for $\D=1-1/q$ and $\a=h+1/q$ as given in \cite{Das:2020kmt,Jevicki:2016bwu}. Thus the full results for the integral in \eqref{kerg} can be written as
\bal
s(\t_1,\t_2)=&\int d\t_3d\t_4\; \ck(\t_1 , \t_2 ; \t_3 , \t_4 )\;g(\t_3,\t_4),\nn\\
=&\left(\;2^{q-1} b^{q-2}\cB\;\(1+i\sqrt{\frac{2\l(1+\l)}{N^{q-1}}}\;\tilde{M}\)\; \frac{\cJ^2 (q-1)}{q(1+\l)}\;\g(h,q)\;\frac{\sgn{(\t_1-\t_2)}}{|\t_1-\t_2|^{2-\frac{2}{q}+2h}}\right),\label{Qsource}
\eal
where 
\bal
\g(h,q)=1-\frac{\pi \Gamma\(\frac{2}{q}\) \csc{\[\pi  \(h+\frac{1}{q}\)\]} \sec{\left[\pi  \(h-\frac{1}{q}\)\]}}{q\; \Gamma \(3-\frac{2}{q}\) \Gamma \(2h+\frac{2}{q}\) \Gamma \(\frac{2}{q}-2h-1\)},\label{gammafunc}
\eal
and it may be noted that the function $\g(h,q)$ vanishes for $h=1/2$. Expanding the function $\g(h,q)$ around $h=1/2$ one can approximate it as 
\bal
\g(h,q)=\frac{6q\;\bar{\g}}{(q-1)b^{q-1}}\(h-\frac{1}{2}\)+\cO\(\(h-\frac{1}{2}\)^2\),\quad \bar{\g}=\frac{(q-1)b^{q-1}}{6q} \g'\(h=\frac{1}{2},q\),\label{gammaExpand}
\eal
where $\g'$ is the derivative of $\g$ function in \eqref{gammafunc} with respect to h. Thus expanding the right hand side of \eqref{Qsource} around $h=1/2$ and using \eqref{gammaExpand}, the source $s(\t_1,\t_2)$ can now be given as
\bal
s(\t_1,\t_2)\simeq &\left(2^{q-1} 6\;q \;b^{-1}\cB\;\bar{\g}\;\(1+i\sqrt{\frac{2\l(1+\l)}{N^{q-1}}}\;\tilde{M}\)\; \frac{\cJ^2}{q(1+\l)}\frac{\sgn{(\t_1-\t_2)}}{|\t_1-\t_2|^{2-\frac{2}{q}+2h}}\right)\(h-\frac{1}{2}\)\nn\\
&+\cO\(\(h-\frac{1}{2}\)^2\),\label{QsourceS}
\eal
which implies that in the limit $h\to 1/2$, the source vanishes and the condition in \eqref{kerg} becomes homogeneous. In this sense, the source $s(\t_1,\t_2)$ is said to be off-shell regularized slightly away from the $h=1/2$ value. 

Plugging back the form of the source \eqref{QsourceS} obtained above in the action \eqref{geff} and then taking the limit $h\to 1/2$, one can see that the second and fourth terms vanish trivially in its expression. In this limit, the action $S$ can now be given as
\bal
\frac{S}{N}=&\cS_{c}[G_c]-\frac{1}{2}\lim_{h\to \frac{1}{2}}\int d\t_1d\t_2\; G_c(\t_1,\t_2)\left(2^{q-1} 6\;q \;b^{-1}\cB\;\bar{\g}\;\(1+i\sqrt{\frac{2\l(1+\l)}{N^{q-1}}}\;\tilde{M}\)\; \right.\nn\\
&\left.\frac{\cJ^2}{q(1+\l)}\;\frac{\sgn{(\t_1-\t_2)}}{|\t_1-\t_2|^{2-\frac{2}{q}+2h}}\(h-\frac{1}{2}\)+\cO\(\(h-\frac{1}{2}\)^2\)\right),\nn\\
\simeq &\;\cS_{c}[G_c]-2^{q-1} 6\;q \;b^{-1}\cB\;\bar{\g}\;\(1+i\sqrt{\frac{2\l(1+\l)}{N^{q-1}}}\;\tilde{M}\)\frac{\cJ^2}{q(1+\l)}\nn\\
&~~~~~~~~~~~~~\(\lim_{h\to \frac{1}{2}}\(h-\frac{1}{2}\)\int d\t_1d\t_2\;\frac{\sgn{(\t_1-\t_2)}}{|\t_1-\t_2|^{2+2h}}\),\label{schAction}
\eal
where the only nontrivial contribution comes from the second term in the above expression. As elucidated earlier, the above action remains invariant under the transformation $\t\to\phi(\t)$ where $\phi(\t)\in SL(2, R)$. These symmetry modes can be uplifted to become dynamical variables using the Faddeev-Popov method suggested in \cite{Das:2020kmt,Jevicki:2016bwu} which at the end gives us the following partition function corresponding to the $\phi(\t)$ modes
\bal
\cZ(\l,\cJ)=\frac{1}{\sqrt{2\pi}}\int\limits_{-\infty}^{\infty} d\tilde{M} \exp{\[-\frac{\tilde{M}^2}{2}-\frac{S}{N}\]},\label{ZschAct}
\eal
where 
\bal
\frac{S}{N}=&\;\cS_{c}[G^{\phi}_c]-2^{q-1} 6\;q \;b^{-1}\cB\;\bar{\g}\;\(1+i\sqrt{\frac{2\l(1+\l)}{N^{q-1}}}\;\tilde{M}\)\frac{\cJ^2}{q(1+\l)}\nn\\
&\(\lim_{h\to \frac{1}{2}}\(h-\frac{1}{2}\)\int d\t_1d\t_2\;\(\frac{\sqrt{\abs{\phi(\t_1)\phi(\t_2)}}}{|\phi(\t_1)-\phi(\t_2)|}\)^{2+2h}\sgn{(\phi(\t_1)-\phi(\t_2))}\),\label{schAction1}
\eal
for which we have made a change to the integration variable from $\t \to \phi(\t)$ in the action \eqref{schAction}. It was shown in \cite{Das:2020kmt,Jevicki:2016bwu} that the second term in \eqref{schAction1} actually gives the Schwarzian action as the integral under parenthesis has a simple pole at  $h=1/2$ which exactly cancels out with the $(h-1/2)$ factor present in the source $s(\t_1,\t_2)$ giving a finite result. Thus following \cite{Das:2020kmt,Jevicki:2016bwu} we have
\bal
\frac{S}{N}=&\;\cS_{c}[G^{\phi}_c]-\frac{\cB\bar{\g}}{2\sqrt{\(1+i\sqrt{\frac{2\l(1+\l)}{N^{q-1}}}\;\tilde{M}\)\frac{2^{q-1}\cJ^2}{ q(1+\l)}}}\int d\t\{\phi(\t),\t\},\label{schAct}
\eal
where the second term
\bal
\int d\t\{\phi(\t),\t\}=\int d\t \(\frac{\phi'''(\t)}{\phi'(\t)}-\frac{3}{2}\(\frac{\phi''(\t)}{\phi'(\t)}\)^2\),\label{sch}
\eal
in the above expression is the zero temperature Schwarzian action and the prime denotes the derivative with respect to the $\t$-coordinate. For a small but finite temperature case, using the following transformation
\bal
\phi(\t)\to\tan{\(\frac{\pi \phi(\t)}{\b}\)},\label{fintemp}
\eal
one can also obtain the finite temperature version of the action in \eqref{schAct} as
\bal
\frac{S}{N}=&\;\cS_{c}[G^{\phi}_c]-\frac{\cB\bar{\g}}{2\sqrt{\(1+i\sqrt{\frac{2\l(1+\l)}{N^{q-1}}}\;\tilde{M}\)\frac{2^{q-1}\cJ^2}{ q(1+\l)}}}\(\int\limits_{0}^{\b} d\t\left\{\tan{\[\frac{\pi \phi(\t)}{\b}\]},\t\right\}\).\label{schActFT}
\eal

In order to determine the Schwarzian action for the SYK model with non-Gaussian disorder we evaluate the partition function in \eqref{ZschAct} for the action \eqref{schAct} given above,
\bal
\cZ(\l,\cJ)=&\frac{1}{\sqrt{2\pi}}\int\limits_{-\infty}^{\infty} d\tilde{M} \exp{\[-\frac{\tilde{M}^2}{2}-\frac{S}{N}+\cS_{c}[G^{\phi}_c]\]},\nn\\
=&\frac{1}{\sqrt{2\pi}}\int\limits_{-\infty}^{\infty} d\tilde{M} \exp\left[-\frac{\tilde{M}^2}{2}+\frac{\cB\;\bar{\g}\sqrt{\;q (1+\l)}}{\cJ \sqrt{2^{q+1}}}\(1-i\sqrt{\frac{\l(1+\l)}{2N^{q-1}}}\;\tilde{M}\)\right.\nn\\
&~~~~~~~~~~~~~~~~~~~~~~~~~~~~~\left.\int d\t\{\phi(\t),\t\}+\cO\(\frac{1}{N^{q-1}}\)\right],\label{Sactsch}
\eal
where in the second line we have made a large $N$ expansion in order to approximate the integral. Thus on integrating over the auxiliary field $\tilde{M}$ one arrives at the following effective action
\bal
\frac{\cI_{sch}}{N}=&-\frac{\cB\;\bar{\g}\sqrt{\;q (1+\l)}}{\cJ \sqrt{2^{q+1}}}\int d\t\{\phi(\t),\t\}\nn\\
&+\frac{\l\;q\;\cB^2\;\bar{\g}^2(1+\l)^2}{2^{q+3}\cJ^2 N^{q-1}}\int \int  d\t d\t'\{\phi(\t),\t\}\{\phi(\t'),\t'\},\label{SactschNG}
\eal
for the non-Gaussian SYK model in the zero temperature regime. However, in the near-infrared (small but finite temperature) regime, one can use the transformation in \eqref{fintemp} to obtain the finite temperature version of the above action.

\subsection{Out of time order correlators (OTOC) and Chaos}
An important aspect of the Schwarzian theory is its maximally chaotic behavior \cite{Maldacena:2016hyu} which is usually captured by the out-of-time order correlation function. To see this we first focus on the following partition function for the Gaussian SYK (GS) model in the near-infrared regime
\bal
\cZ^{GS}=&\int \frac{d\mu(\phi)}{SL(2,R)} \;\exp\left[\frac{\cB\;\bar{\g}\sqrt{q}}{\cJ \sqrt{2^{q+1}}}\int d\t\left\{\tan{\[\frac{\pi \phi(\t)}{\b}\]},\t\right\}\right],\label{SchGSZ}
\eal
where $d\mu(\phi)$ is the appropriate measure for the reparametrization modes $\phi(\t)\in SL(2, R)$. The effective action corresponding to the above partition function can now be given as
\bal
\frac{\cI^{GS}}{N}&=-\frac{\cB\;\bar{\g}\sqrt{q }}{\cJ \sqrt{2^{q+1}}}\int\limits_{0}^{\b} d\t\(\left\{\phi(\t),\t\right\}+\frac{2 \pi ^2 \phi '(\tau )^2}{\beta ^2}\),\label{IschGS}
\eal
which is obtained by first taking the transformation \eqref{fintemp} and then setting $\l=0$ in \eqref{SactschNG}. The mode $\phi(\t)$ satisfies the periodic boundary conditions $\phi(\t+\b)=\phi(\t)$. The periodic boundary condition helps us to write the Schwarzian action in \eqref{IschGS} as
\bal
\frac{\cI^{GS}}{N}&=-\frac{\cB\;\bar{\g}\sqrt{q}}{\cJ \sqrt{2^{q+1}}}\int\limits_{0}^{\b} d\t\(-\frac{1}{2}\(\frac{\phi''(\t)}{\phi'(\t)}\)^2+\frac{2 \pi ^2 \phi '(\tau )^2}{\beta ^2}\),\label{IschGS1}
\eal

We now use the following expansion 
\bal
\phi(\t)=\t+\epsilon(\t),\quad \epsilon(\t+\b)=\epsilon(\t),\label{epstrans}
\eal
which gives the action in \eqref{IschGS1} to the leading order in $\epsilon(\t)$ as
\bal
\frac{\cI_{\epsilon}^{GS}}{N}&=-\frac{\cB\;\bar{\g}\sqrt{q}}{\cJ \sqrt{2^{q+1}}}\int\limits_{0}^{\b} d\t\(\frac{4 \pi ^2 \epsilon'(\tau )^2}{\beta ^2}-\epsilon''(\t)^2+\cO(\epsilon^3)\),\label{IschGS12}
\eal
where in writing above we have dropped off a constant term appearing in the above action. Rescaling the time coordinate as $\tau \to 2\pi \tau/\b$, the above effective action can be rewritten as
\bal
\frac{\cI_{\epsilon}^{GS}}{N}&=-\frac{8\pi^3}{\b^3}\frac{\cB\;\bar{\g}\sqrt{q}}{\cJ \sqrt{2^{q+1}}}\int\limits_{0}^{2\pi} d\t\( \epsilon'(\tau )^2-\epsilon''(\t)^2\),\label{IschGS2}
\eal
whose corresponding equation of motion can be given by
\bal
\epsilon^{(4)}(\t)+\epsilon''(\tau )=0.\label{IschGSeom}
\eal

The solutions to the above equation of motion for the periodic boundary condition in \eqref{epstrans} are given by $\epsilon(\t)=e^{i  n \t}$, with $n=\left\{0,\pm 1\right\}$.  Moreover, the two-point function for the $\epsilon(\t)$ field can now be obtained as
\bal
G^{GS}_{\epsilon}(\t,0)=\VEV{\epsilon(\t)\epsilon(0)}_{GS}=\frac{\b^3}{8\pi^3}\(\frac{\cB\;\bar{\g}\sqrt{q}}{\cJ \sqrt{2^{q+1}}}\)^{-1}\sum\limits_{\abs{n}\geq 2}\frac{e^{i n\t}}{n^2(n^2-1)},\label{IschGSEP}
\eal
which has poles at $n=\left\{0,\pm 1\right\}$, corresponding to the unbroken $SL(2, R)$ gauge symmetry and so we should not sum over these modes. The sum in the above propagator can be evaluated as a contour integral where the contour is taken such that it only encircles the poles at $n=\left\{0,\pm 1\right\}$ as shown in the figure. This gives us a Matsubara integral for the frequencies $\om=i n$, as follows
\bal
\VEV{\epsilon(\t)\epsilon(0)}_{GS}=\frac{\b^3}{8\pi^3}\(\frac{\cB\;\bar{\g}\sqrt{q}}{\cJ \sqrt{2^{q+1}}}\)^{-1}\int\limits_{\cC}d\om\;\frac{e^{\om \t}}{e^{2\pi\om}-1}\(\frac{1}{\om^2(\om^2+1)}\),\label{IschGSEP1}
\eal
which in turn can be evaluated as a sum of contour integrals around the poles $\om=\left\{0,\pm i\right\}$ as 
\bal
\VEV{\epsilon(\t)\epsilon(0)}_{GS}=&\frac{\b^3}{16\pi^3} \(\frac{\cB\;\bar{\g}\sqrt{q}}{\cJ \sqrt{2^{q+1}}}\)^{-1}\left(-5 \cos \left(\frac{2 \pi  \tau }{\beta }\right)+\frac{4 \pi ^2 \tau ^2}{\beta ^2}\right.\nn\\
&\left.+\left(2\pi  \sin \left(\frac{2 \pi  \tau }{\beta }\right)+\frac{2\pi ^2}{3}-2\right)-\frac{4 \left( \pi  \tau  \sin \left(\frac{2 \pi  \tau }{\beta }\right)+ \pi ^2 \tau \right)}{\beta }\).\label{IschGSEP2}
\eal

Having determined the $\epsilon$-propagator, we now move to evaluate the two and four-point functions for the operators $V(\t)$ and $W(\t)$ with conformal dimensions $\D_{v}$ and $\D_{w}$ which are made up of the product of
Majorana fermion bilinears for the usual Gaussian SYK model. First, we consider the thermal two-point function for the field $V(\phi)$ which can be obtained from \eqref{conGb} as
\bal
\VEV{V(\t_1)V(\t_2)}_{GS}&=b\[\frac{\pi}{\b\sin\(\frac{\pi\t_{12}}{\b}\)}\]^{2\D_{V}},
\eal
then on using the transformation $\t\to \phi(\t)$, we get
\bal
\VEV{V(\phi_1)V(\phi_2)}_{GS}&=b\[\frac{\pi\sqrt{\phi'(\t_1)\phi'(\t_2)}}{\b\sin\(\frac{\pi\(\phi(\t_1)-\phi(\t_2)\)}{\b}\)}\]^{2\D_{V}},\quad \b>\t_{12} >0\nn\\
&\simeq \VEV{V(\t_1)V(\t_2)}_{GS}\(1+\frac{\d\VEV{V(\t_1)V(\t_2)}_{GS}}{\VEV{V(\t_1)V(\t_2)}_{GS}}\),\label{2pGSV}
\eal
where $\phi_i=\phi(\t_i)$, $\t_{ij}=(\t_i-\t_j)$, and
\bal
\frac{\d\VEV{V(\t_1)V(\t_2)}_{GS}}{\VEV{V(\t_1)V(\t_2)}_{GS}}=\D_{V}\(\epsilon'(\t_1)+\epsilon'(\t_2)-\frac{2\pi}{\b}\;\(\frac{\epsilon(\t_1)-\epsilon(\t_2)}{\tan\(\frac{\pi\t_{12}}{\b}\)}\)\).
\eal

In the second line of \eqref{2pGSV} considering $\phi(\t)=\t+\epsilon(\t)$, we have expanded the two-point function to the linear order in $\epsilon$. In a similar way as above, one can also obtain the thermal two-point function for the $W(\phi)$ field. In the two-point function for the field $V$, we now substitute $\epsilon(\t)=e^{i \frac{2\pi n \t}{\b}}$ which gives us
\bal
\frac{\d\VEV{V(\t_1)V(\t_2)}_{GS}}{\VEV{V(\t_1)V(\t_2)}_{GS}}&=i\frac{4\pi\D_{V}}{\b}\sgn{(n)}\;e^{i \frac{\pi n (\t_1+\t_2)}{\b}}\(\abs{n} \cos{\(\frac{\abs{n}\pi\t_{12}}{\b}\)}-\frac{\sin{\(\frac{\abs{n}\pi\t_{12}}{\b}\)}}{\tan{\(\frac{\pi\t_{12}}{\b}\)}}\),\label{2pGSV1delta}
\eal
which vanishes for the modes corresponding to $n=\left\{0,\pm 1\right\}$. As we will see later that this relation will be important in determining the four-point function for the $V$ and $W$ fields.

We next consider the  four-point function for the fields $V$ and $W$ which in the large N limit can be given as
\bal
&\VEV{V(\t_1)V(\t_2)W(\t_3)W(\t_4)}_{GS}\simeq \nn\\&\VEV{V(\t_1)V(\t_2)}_{GS}\VEV{W(\t_3)W(\t_4)}_{GS}
\begin{cases}
  \(1+N \cF^{TO}(\t_1,\t_2,\t_3,\t_4)\), & 0\leq\t_1<\t_2<\t_3<\t_4\leq\b \\
 \pm  \(1+N \cF^{OTO}(\t_1,\t_2,\t_3,\t_4)\), & 0\leq\t_1<\t_3<\t_2<\t_4\leq\b
\end{cases},\label{ZT4pVWGS}
\eal
where the superscript $TO$ stands for the time-ordered four-point function and the superscript $OTO$ stands for the out-of-time ordered four-point function. The leading correction to the four-point function $\cF$ denotes the connected part of the four-point function \cite{Maldacena:2016hyu, Kitaev:2017awl}. In the time-ordered case the function $\cF^{TO}$ can be given as
\bal
\cF^{TO}(\t_1,\t_2,\t_3,\t_4)&=\VEV{\(\frac{\d\VEV{V(\t_1)V(\t_2)}_{GS}}{\VEV{V(\t_1)V(\t_2)}_{GS}}\)\(\frac{\d\VEV{W(\t_3)W(\t_3)}_{GS}}{\VEV{W(\t_3)W(\t_4)}_{GS}}\)},\nn\\
&=\frac{2\D_{V}\D_{W}}{\pi}\(\frac{\cB\;\bar{\g}\sqrt{q}}{\cJ \sqrt{2^{q+1}}}\)^{-1}\sum\limits_{\abs{n}\geq 2}\frac{e^{i\frac{\pi n \x_{+}}{\b}}}{n^2(n^2-1)} \(\abs{n} \cos{\(\frac{\abs{n}\pi\x}{\b}\)}-\frac{\sin{\(\frac{\abs{n}\pi\x}{\b}\)}}{\tan{\(\frac{\pi\x}{\b}\)}}\)\nn\\
&\(\abs{n} \cos{\(\frac{\abs{n}\pi\x'}{\b}\)}-\frac{\sin{\(\frac{\abs{n}\pi\x'}{\b}\)}}{\tan{\(\frac{\pi\x'}{\b}\)}}\),
\eal
where
\bal
\x_{+}=\(\t_{1}+\t_{2}\)-\(\t_{3}+\t_{4}\),\quad \x=\t_{12},\quad \x'=\t_{34}
\eal

Now in order to evaluate the sum in the expression for $\cF^{TO}$ given above, we convert it to a Matsubara contour integral as we did in order to evaluate the $\epsilon$ propagator in \eqref{IschGSEP1}. Then evaluating the subsequent integral as a sum of contour integrals around the poles $\om=\left\{0,\pm i\right\}$ we obtain the result for the function $\cF^{TO}$ as follows
\bal
\cF^{TO}(\t_1,\t_2,\t_3,\t_4)&=\frac{\D_{V}\D_{W}}{\pi^2}\(\frac{\cB\;\bar{\g}\sqrt{q}}{\cJ \sqrt{2^{q+1}}}\)^{-1}\left(1-\frac{\pi\;\x }{\beta \tan\left(\frac{\pi\;\x}{\beta }\right)}\right)\left(1-\frac{\pi\;\x'}{\beta  \tan \left(\frac{\pi\;\x'}{\beta }\right) }\right)
\eal

In the case of the out-of-time order four-point function, determining $\cF^{OTO}$ is in general complicated. However, the situation simplifies in the case when fields $V$ or $W$ belonging to one of the pairs are situated at the antipodal points of the thermal circle. Specifically, we make the following choice for the time arguments
\bal
\t_3 \to 0,\quad \t_4 \to \frac{\b}{2}
\eal
which gives 
\bal
\cF^{OTO}(\t_1,\t_2,\t_3,\t_4)&=\frac{2\D_{V}\D_{W}}{\pi}\(\frac{\cB\;\bar{\g}\sqrt{q}}{\cJ \sqrt{2^{q+1}}}\)^{-1}\sum\limits_{\abs{n}\geq 2}\frac{e^{i\frac{2\pi n \x_{+}}{\b}}}{n^2(n^2-1)} \(\abs{n} \cos{\(\frac{\abs{n}\pi\x}{\b}\)}-\frac{\sin{\(\frac{\abs{n}\pi\x}{\b}\)}}{\tan{\(\frac{\pi\x}{\b}\)}}\)\nn\\
&\(\abs{n} \cos{\(\frac{\abs{n}\pi\x'}{\b}\)}-\frac{\sin{\(\frac{\abs{n}\pi\x'}{\b}\)}}{\tan{\(\frac{\pi\x'}{\b}\)}}\),\nn\\
&=\frac{\D_{V}\D_{W}}{\pi^2}\(\frac{\cB\;\bar{\g}\sqrt{q}}{\cJ \sqrt{2^{q+1}}}\)^{-1}\left(1+\frac{\b}{2} \frac{\sin\(\frac{\pi\t_1}{\b}\)\sin\(\frac{\pi\t_2}{\b}\)}{\abs{\sin\(\frac{\pi\x}{\b}\)}}-\frac{\pi\;\x }{\beta \tan\left(\frac{\pi\;\x}{\beta }\right)}\right)
\eal
such that on going to Lorentzian time for $\t_1=\t_2-\frac{\b}{2}$ and $\t_2=\frac{\b}{4}-i t$ we have
\bal
\cF^{OTO}(\t_1,\t_2,\t_3,\t_4)&=\frac{\D_{V}\D_{W}}{\pi^2}\(\frac{\cB\;\bar{\g}\sqrt{q}}{\cJ \sqrt{2^{q+1}}}\)^{-1}\(1-\frac{\b}{4}\cosh\(\l_{l} t\)\)
\eal
where $\l_{l}=\frac{2\pi}{\b}$, is the Lyapunov exponent which saturates the chaos bound \cite{Maldacena:2015waa}. To this end, it should be noted that the computation of the time ordered or the out-of-time-ordered four-point function depends on the $\epsilon$-propagator \eqref{IschGSEP2} which has a piece that grows as $e^{2\pi t/ \b}$ at late Lorentzian ($\t\to it $) times giving the maximal chaotic behavior for the Gaussian SYK model. We now move on to determine this propagator for the non-Gaussian SYK model. For this, we first consider the effective action for the partition function in \eqref{Sactsch} as
\bal
\cI(\tilde{M})=\frac{\tilde{M}^2}{2}-\frac{\cB\;\bar{\g}\sqrt{\;q (1+\l)}}{\cJ \sqrt{2^{q+1}}}\(1-i\sqrt{\frac{\l(1+\l)}{2N^{q-1}}}\;\tilde{M}\)\int d\t\{\phi(\t),\t\},\label{IMschNGS2}
\eal 
which on using the transformation \eqref{epstrans} and expanding to the second order in $\epsilon$ can be written as
\bal
\cI_{\epsilon}(\tilde{M})=\frac{\tilde{M}^2}{2}-\frac{\cB\;\bar{\g}\sqrt{\;q (1+\l)}}{\cJ \sqrt{2^{q+1}}}\(1-i\sqrt{\frac{\l(1+\l)}{2N^{q-1}}}\;\tilde{M}\)\int\limits_{0}^{\b} d\t\(\frac{4 \pi ^2 \epsilon'(\tau )^2}{\beta ^2}-\epsilon''(\t)^2\).\label{IMschNGS1}
\eal

The partition function for the above effective action corresponding to the non-Gaussian SYK model can be given as
\bal
\cZ_{\epsilon}(\cQ)=&\frac{1}{\sqrt{2\pi}}\int d\epsilon(\t)\int\limits_{-\infty}^{\infty} d\tilde{M} \;\exp\left[-\cI_{\epsilon}\(\tilde{M}\)-\int d\t\;\cQ(\t)\;\epsilon(\t) \right],\label{SchNGSZ}
\eal
where we have added a term involving the nondynamical source $\cQ(\t)$ to the effective action in \eqref{IMschNGS1} given before. Adding such a source term is a usual step in calculating the n-point correlation functions which are given as $n^{th}$ functional derivatives of the partition function with respect to the source where one sets $\cQ \to 0$, at the end. Next using the following transformation
\bal
\epsilon(\t)=\frac{\bar{\epsilon}(\t)}{\sqrt{\sqrt{(1+\l)}\(1-i\sqrt{\frac{\l(1+\l)}{2N^{q-1}}}\;\tilde{M}\)}},\label{epstransM}
\eal
one gets the partition function in \eqref{SchNGSZ} as
\bal
\cZ_{\epsilon}(\cQ)=&\frac{1}{\sqrt{2\pi}}\int\limits_{-\infty}^{\infty} d\tilde{M}\;\int \frac{d\epsilon(\t)}{\sqrt{\sqrt{(1+\l)}\(1-i\sqrt{\frac{\l(1+\l)}{2N^{q-1}}}\;\tilde{M}\)}}\;\exp\left[-\frac{\tilde{M}^2}{2}\right.\nn\\
&\left.-\frac{\cB\;\bar{\g}\sqrt{q}}{\cJ \sqrt{2^{q+1}}}\int\limits_{0}^{\b} d\t\;\epsilon(\t)\(\pa^{(4)}_{\t}+\frac{4 \pi ^2 }{\beta ^2}\pa''_{\t}\)\epsilon(\t)-\int d\t\;\frac{\cQ(\t)\;\epsilon(\t)}{\sqrt{\sqrt{(1+\l)}\(1-i\sqrt{\frac{\l(1+\l)}{2N^{q-1}}}\;\tilde{M}\)}}\right],\label{SchNGSZM}
\eal
where getting the above result after the transformation, we have removed the bar over the $\epsilon$ field as it is only an integration variable. Integrating over the $\epsilon(\t)$ we get the following result for the above partition function
\bal
\cZ_{\epsilon}(\cQ)=&\frac{\cZ_{\epsilon}(0)}{\sqrt{2\pi}}\int\limits_{-\infty}^{\infty} \frac{d\tilde{M}}{\sqrt{\sqrt{ (1+\l)}\(1-i\sqrt{\frac{\l(1+\l)}{2N^{q-1}}}\;\tilde{M}\)}}\;\exp\left[-\frac{\tilde{M}^2}{2}\right.\nn\\
&\left.+\int d\t\;d\t'\;\frac{\cQ(\t)\;G^{GS}_{\epsilon}(\t,\t')\;\cQ(\t')}{2\sqrt{(1+\l)}\(1-i\sqrt{\frac{\l(1+\l)}{2N^{q-1}}}\;\tilde{M}\)}\right],\label{SchNGSZM1}
\eal
where 
\bal
\cZ_{\epsilon}(0)=\frac{1}{\sqrt{\frac{\cB\;\bar{\g}\sqrt{q}}{\cJ \sqrt{2^{q+1}}}\det\[\pa^{(4)}_{\t}+\frac{4 \pi ^2 }{\beta ^2}\pa''_{\t}\]}},\label{SchNGSZ0}
\eal
and $G^{GS}_{\epsilon}(\t,\t')$ is the $\epsilon$-propagator defined in \eqref{IschGSEP} for the Gaussian SYK model. We now expand the partition function in \eqref{SchNGSZM1} the large N limit which gives us
\bal
\cZ_{\epsilon}(\cQ)\simeq &\frac{\cZ_{\epsilon}(0)}{\sqrt{2\pi}}\int\limits_{-\infty}^{\infty} d\tilde{M}\;\exp\left[-\frac{\tilde{M}^2}{2}+i\sqrt{\frac{\l(1+\l)}{8N^{q-1}}}\;\tilde{M}-\frac{1}{4}\log{\(1+\l\)}\right.\nn\\
&\left.+\frac{1}{2\sqrt{(1+\l)}}\(1+i\sqrt{\frac{\l(1+\l)}{2N^{q-1}}}\;\tilde{M}\)\int d\t\;d\t'\;\cQ(\t)\;G^{GS}_{\epsilon}(\t,\t')\;\cQ(\t')+\cO\(\frac{1}{N^{q-1}}\)\right],\label{SchNGSZM2}
\eal

One can now determine the $\epsilon$-propagator $G^{NGS}_{\epsilon}$ for the non-Gaussian SYK model as
\bal
G^{NGS}_{\epsilon}(\t,\t')=&\left.\frac{\d^2 \cZ_{\epsilon}(\cQ)}{\d\cQ(\t)\d\cQ(\t')}\right|_{\cQ=0},\nn\\
=&\frac{1}{\sqrt{2\pi}}\int\limits_{-\infty}^{\infty} d\tilde{M}\;\frac{1}{\sqrt{(1+\l)}}\(1+i\sqrt{\frac{\l(1+\l)}{2N^{q-1}}}\;\tilde{M}\)\;G^{GS}_{\epsilon}(\t,\t')\nn\\
&\exp\left[-\frac{\tilde{M}^2}{2}+i\sqrt{\frac{\l(1+\l)}{8N^{q-1}}}\;\tilde{M}-\frac{1}{4}\log{\(1+\l\)}\right],\nn\\
=&\; G^{GS}_{\epsilon}(\t,\t')\(\frac{1}{(\lambda +1)^{3/4}}-\frac{3\l\(1+\l\)^{1/4}}{16\;N^{q-1}}\)+\cO\(\frac{1}{N^{q-1}}\),\label{SchNGSEP}
\eal
where in obtaining the last line in the above expression we have first integrated over the auxiliary field $\tilde{M}$ and then performed a large N expansion. From the form of the $\epsilon$-propagator for the non-Gaussian SYK model, it is obvious to conclude that the Lyapunov exponent will remain unchanged as the four-point OTOC depends on the late Lorentzian time behavior of this propagator.

\section{Summary and Discussion}
 The study of the effects of non-Gaussian quenched disorder on the SYK model with Majorana fermions within the framework of large N expansion reveals intriguing insights into the behavior of this complex system. The non-Gaussian disorder averaging induces modification of the variance of the Gaussian distribution of couplings to the leading order in N. This also gives rise to a subdominant non-local term in the large N effective action. These findings are also observed in the case of the SYK model with flavored complex fermions, suggesting a broader applicability of introducing non-Gaussian disorder. Exploring the physical quantities in the low-temperature regime emphasizes that, despite its non-Gaussian nature, the model retains considerable similarity to the Gaussian SYK model at leading order in N. A particularly notable result is the insensitivity of the Lyapunov exponent to non-Gaussianity, underscoring the model's robust chaotic behavior. In the large N and large q limit one uncovers a connection between the non-Gaussian SYK model and a non-local Liouville field theory.  This invokes intriguing parallels to ideas proposed by Veneziano and Coleman in the context of nonlocal Euclidean quantum gravity theory with wormholes solutions. The presence of wormhole solutions suggests that the cosmological constant can be tuned to be very small which we have demonstrated for the non-local Liouville field theory dual to the non-Gaussian SYK model. Overall, this investigation demonstrates the intricate interplay between non-Gaussian SYK model, two-dimensional non local Euclidean quantum gravity, and its wormhole solutions which further expands our comprehension of such complex quantum systems and their gravity duals.

 One can also investigate several future directions. First, one may consider investigating the Coleman's mechanism in the case of two coupled SYK model discussed in \cite{Maldacena:2018lmt}. It would also be interesting to explore the same mechanism for a large number of coupled SYK models that might describe a two-dimensional multiverse. One can then use this multiverse model to explain the observed small cosmological constant. Finally, our 2D gravity action is classical. In \cite{Goel:2023svz} it was shown that quantum fluctuations in the double scaling limit of SYK modify the geometry. It would be interesting to repeat such a calculation in the non-Gaussian disorder setting we suggested. We leave these interesting avenues for future works.

\section*{Acknowledgement}

The authors would like to thank Soumangsu Chakraborty, Antal Jevicki, and Misha Smolkin for valuable discussions. PC is supported by the postdoctoral program at the Ariel University. 

\bibliographystyle{JHEP}

\bibliography{ref.bib}

\providecommand{\href}[2]{#2}\begingroup\raggedright\begin{thebibliography}{10}

\bibitem{Sachdev:1993PhRvL}
S.~{Sachdev} and J.~{Ye}, \emph{{Gapless spin-fluid ground state in a random quantum Heisenberg magnet}}, \href{https://doi.org/10.1103/PhysRevLett.70.3339}{\emph{Physical Review Letters} {\bfseries 70} (1993) 3339} [\href{https://arxiv.org/abs/cond-mat/9212030}{{\ttfamily cond-mat/9212030}}].

\bibitem{Sachdev:2010uj}
S.~Sachdev, \emph{{Strange metals and the AdS/CFT correspondence}}, \href{https://doi.org/10.1088/1742-5468/2010/11/P11022}{\emph{J. Stat. Mech.} {\bfseries 1011} (2010) P11022} [\href{https://arxiv.org/abs/1010.0682}{{\ttfamily 1010.0682}}].

\bibitem{Parcollet:1999oa}
O.~{Parcollet} and A.~{Georges}, \emph{{Non-Fermi-liquid regime of a doped Mott insulator}}, \href{https://doi.org/10.1103/PhysRevB.59.5341}{\emph{Phys. Rev. B} {\bfseries 59} (1999) 5341} [\href{https://arxiv.org/abs/cond-mat/9806119}{{\ttfamily cond-mat/9806119}}].

\bibitem{Davison:2016ngz}
R.A.~Davison, W.~Fu, A.~Georges, Y.~Gu, K.~Jensen and S.~Sachdev, \emph{{Thermoelectric transport in disordered metals without quasiparticles: The Sachdev-Ye-Kitaev models and holography}}, \href{https://doi.org/10.1103/PhysRevB.95.155131}{\emph{Phys. Rev.} {\bfseries B95} (2017) 155131} [\href{https://arxiv.org/abs/1612.00849}{{\ttfamily 1612.00849}}].

\bibitem{Gu:2016oyy}
Y.~Gu, X.-L.~Qi and D.~Stanford, \emph{{Local criticality, diffusion and chaos in generalized Sachdev-Ye-Kitaev models}}, \href{https://doi.org/10.1007/JHEP05(2017)125}{\emph{JHEP} {\bfseries 05} (2017) 125} [\href{https://arxiv.org/abs/1609.07832}{{\ttfamily 1609.07832}}].

\bibitem{Witten:2016iux}
E.~Witten, \emph{{An SYK-Like Model Without Disorder}},  \href{https://arxiv.org/abs/1610.09758}{{\ttfamily 1610.09758}}.

\bibitem{Fu:2016vas}
W.~Fu, D.~Gaiotto, J.~Maldacena and S.~Sachdev, \emph{{Supersymmetric Sachdev-Ye-Kitaev models}}, \href{https://doi.org/10.1103/PhysRevD.95.069904, 10.1103/PhysRevD.95.026009}{\emph{Phys. Rev.} {\bfseries D95} (2017) 026009} [\href{https://arxiv.org/abs/1610.08917}{{\ttfamily 1610.08917}}].

\bibitem{Klebanov:2016xxf}
I.R.~Klebanov and G.~Tarnopolsky, \emph{{Uncolored random tensors, melon diagrams, and the Sachdev-Ye-Kitaev models}}, \href{https://doi.org/10.1103/PhysRevD.95.046004}{\emph{Phys. Rev.} {\bfseries D95} (2017) 046004} [\href{https://arxiv.org/abs/1611.08915}{{\ttfamily 1611.08915}}].

\bibitem{Gross:2016kjj}
D.J.~Gross and V.~Rosenhaus, \emph{{A Generalization of Sachdev-Ye-Kitaev}}, \href{https://doi.org/10.1007/JHEP02(2017)093}{\emph{JHEP} {\bfseries 02} (2017) 093} [\href{https://arxiv.org/abs/1610.01569}{{\ttfamily 1610.01569}}].

\bibitem{Krishnan:2016bvg}
C.~Krishnan, S.~Sanyal and P.N.~Bala~Subramanian, \emph{{Quantum Chaos and Holographic Tensor Models}}, \href{https://doi.org/10.1007/JHEP03(2017)056}{\emph{JHEP} {\bfseries 03} (2017) 056} [\href{https://arxiv.org/abs/1612.06330}{{\ttfamily 1612.06330}}].

\bibitem{Turiaci:2017zwd}
G.~Turiaci and H.~Verlinde, \emph{{Towards a 2d QFT Analog of the SYK Model}}, \href{https://doi.org/10.1007/JHEP10(2017)167}{\emph{JHEP} {\bfseries 10} (2017) 167} [\href{https://arxiv.org/abs/1701.00528}{{\ttfamily 1701.00528}}].

\bibitem{Murugan:2017eto}
J.~Murugan, D.~Stanford and E.~Witten, \emph{{More on Supersymmetric and 2d Analogs of the SYK Model}}, \href{https://doi.org/10.1007/JHEP08(2017)146}{\emph{JHEP} {\bfseries 08} (2017) 146} [\href{https://arxiv.org/abs/1706.05362}{{\ttfamily 1706.05362}}].

\bibitem{Chen:2017dav}
X.~Chen, R.~Fan, Y.~Chen, H.~Zhai and P.~Zhang, \emph{{Competition between Chaotic and Nonchaotic Phases in a Quadratically Coupled Sachdev-Ye-Kitaev Model}}, \href{https://doi.org/10.1103/PhysRevLett.119.207603}{\emph{Phys. Rev. Lett.} {\bfseries 119} (2017) 207603} [\href{https://arxiv.org/abs/1705.03406}{{\ttfamily 1705.03406}}].

\bibitem{Jian:2017unn}
S.-K.~Jian and H.~Yao, \emph{{Solvable Sachdev-Ye-Kitaev models in higher dimensions: from diffusion to many-body localization}}, \href{https://doi.org/10.1103/PhysRevLett.119.206602}{\emph{Phys. Rev. Lett.} {\bfseries 119} (2017) 206602} [\href{https://arxiv.org/abs/1703.02051}{{\ttfamily 1703.02051}}].

\bibitem{Cai:2017vyk}
W.~Cai, X.-H.~Ge and G.-H.~Yang, \emph{{Diffusion in higher dimensional SYK model with complex fermions}}, \href{https://doi.org/10.1007/JHEP01(2018)076}{\emph{JHEP} {\bfseries 01} (2018) 076} [\href{https://arxiv.org/abs/1711.07903}{{\ttfamily 1711.07903}}].

\bibitem{Peng:2017kro}
C.~Peng, \emph{{Vector models and generalized SYK models}}, \href{https://doi.org/10.1007/JHEP05(2017)129}{\emph{JHEP} {\bfseries 05} (2017) 129} [\href{https://arxiv.org/abs/1704.04223}{{\ttfamily 1704.04223}}].

\bibitem{Maldacena:2018lmt}
J.~Maldacena and X.-L.~Qi, \emph{{Eternal traversable wormhole}},  \href{https://arxiv.org/abs/1804.00491}{{\ttfamily 1804.00491}}.

\bibitem{Maldacena:2015waa}
J.~Maldacena, S.H.~Shenker and D.~Stanford, \emph{{A bound on chaos}}, \href{https://doi.org/10.1007/JHEP08(2016)106}{\emph{JHEP} {\bfseries 08} (2016) 106} [\href{https://arxiv.org/abs/1503.01409}{{\ttfamily 1503.01409}}].

\bibitem{Kitaev:2015tk}
A.~{Kitaev}, \emph{{A simple model of quantum holography}}, {\emph{Talks at KITP, April 7, 2015, and May 27, 2015} } [\href{https://arxiv.org/abs/{http://online.kitp.ucsb.edu/online/entangled15/kitaev/, http://online.kitp.ucsb.edu/online/entangled15/kitaev2/}}{{\ttfamily {http://online.kitp.ucsb.edu/online/entangled15/kitaev/, http://online.kitp.ucsb.edu/online/entangled15/kitaev2/}}}].

\bibitem{Maldacena:2016hyu}
J.~Maldacena and D.~Stanford, \emph{{Remarks on the Sachdev-Ye-Kitaev model}}, \href{https://doi.org/10.1103/PhysRevD.94.106002}{\emph{Phys. Rev.} {\bfseries D94} (2016) 106002} [\href{https://arxiv.org/abs/1604.07818}{{\ttfamily 1604.07818}}].

\bibitem{Maldacena:2016upp}
J.~Maldacena, D.~Stanford and Z.~Yang, \emph{{Conformal symmetry and its breaking in two dimensional Nearly Anti-de-Sitter space}}, \href{https://doi.org/10.1093/ptep/ptw124}{\emph{PTEP} {\bfseries 2016} (2016) 12C104} [\href{https://arxiv.org/abs/1606.01857}{{\ttfamily 1606.01857}}].

\bibitem{Polchinski:2016xgd}
J.~Polchinski and V.~Rosenhaus, \emph{{The Spectrum in the Sachdev-Ye-Kitaev Model}}, \href{https://doi.org/10.1007/JHEP04(2016)001}{\emph{JHEP} {\bfseries 04} (2016) 001} [\href{https://arxiv.org/abs/1601.06768}{{\ttfamily 1601.06768}}].

\bibitem{Bagrets:2017pwq}
D.~Bagrets, A.~Altland and A.~Kamenev, \emph{{Power-law out of time order correlation functions in the SYK model}}, \href{https://doi.org/10.1016/j.nuclphysb.2017.06.012}{\emph{Nucl. Phys. B} {\bfseries 921} (2017) 727} [\href{https://arxiv.org/abs/1702.08902}{{\ttfamily 1702.08902}}].

\bibitem{2021Symm...13..599C}
S.~{Choudhury}, \emph{{The Cosmological OTOC: A New Proposal for Quantifying Auto-Correlated Random Non-Chaotic Primordial Fluctuations}}, \href{https://doi.org/10.3390/sym13040599}{\emph{Symmetry} {\bfseries 13} (2021) 599} [\href{https://arxiv.org/abs/2106.01305}{{\ttfamily 2106.01305}}].

\bibitem{Sonner:2017hxc}
J.~Sonner and M.~Vielma, \emph{{Eigenstate thermalization in the Sachdev-Ye-Kitaev model}}, \href{https://doi.org/10.1007/JHEP11(2017)149}{\emph{JHEP} {\bfseries 11} (2017) 149} [\href{https://arxiv.org/abs/1707.08013}{{\ttfamily 1707.08013}}].

\bibitem{Liu:2017kfa}
C.~Liu, X.~Chen and L.~Balents, \emph{{Quantum Entanglement of the Sachdev-Ye-Kitaev Models}}, \href{https://doi.org/10.1103/PhysRevB.97.245126}{\emph{Phys. Rev. B} {\bfseries 97} (2018) 245126} [\href{https://arxiv.org/abs/1709.06259}{{\ttfamily 1709.06259}}].

\bibitem{Huang:2017nox}
Y.~Huang and Y.~Gu, \emph{{Eigenstate entanglement in the Sachdev-Ye-Kitaev model}}, \href{https://doi.org/10.1103/PhysRevD.100.041901}{\emph{Phys. Rev. D} {\bfseries 100} (2017) 041901} [\href{https://arxiv.org/abs/1709.09160}{{\ttfamily 1709.09160}}].

\bibitem{Zhang:2020sci}
P.~Zhang, C.~Liu and X.~Chen, \emph{{Subsystem Rényi entropy of thermal ensembles for SYK-like models}}, \href{https://doi.org/10.21468/SciPostPhys.8.6.094}{\emph{SciPost Phys.} {\bfseries 8} (2020) 094}.

\bibitem{Zhang:2022yaw}
P.~Zhang, \emph{{Quantum entanglement in the Sachdev\textemdash{}Ye\textemdash{}Kitaev model and its generalizations}}, \href{https://doi.org/10.1007/s11467-022-1162-5}{\emph{Front. Phys. (Beijing)} {\bfseries 17} (2022) 43201} [\href{https://arxiv.org/abs/2203.01513}{{\ttfamily 2203.01513}}].

\bibitem{Haldar:2020ymg}
A.~Haldar, S.~Bera and S.~Banerjee, \emph{{R\'enyi entanglement entropy of Fermi and non-Fermi liquids: Sachdev-Ye-Kitaev model and dynamical mean field theories}}, \href{https://doi.org/10.1103/PhysRevResearch.2.033505}{\emph{Phys. Rev. Res.} {\bfseries 2} (2020) 033505} [\href{https://arxiv.org/abs/2004.04751}{{\ttfamily 2004.04751}}].

\bibitem{Sachdev:2010um}
S.~Sachdev, \emph{{Holographic metals and the fractionalized Fermi liquid}}, \href{https://doi.org/10.1103/PhysRevLett.105.151602}{\emph{Phys. Rev. Lett.} {\bfseries 105} (2010) 151602} [\href{https://arxiv.org/abs/1006.3794}{{\ttfamily 1006.3794}}].

\bibitem{Almheiri:2014cka}
A.~Almheiri and J.~Polchinski, \emph{{Models of AdS$_{2}$ backreaction and holography}}, \href{https://doi.org/10.1007/JHEP11(2015)014}{\emph{JHEP} {\bfseries 11} (2015) 014} [\href{https://arxiv.org/abs/1402.6334}{{\ttfamily 1402.6334}}].

\bibitem{Sachdev:2015efa}
S.~Sachdev, \emph{{Bekenstein-Hawking Entropy and Strange Metals}}, \href{https://doi.org/10.1103/PhysRevX.5.041025}{\emph{Phys. Rev.} {\bfseries X5} (2015) 041025} [\href{https://arxiv.org/abs/1506.05111}{{\ttfamily 1506.05111}}].

\bibitem{Engelsoy:2016xyb}
J.~Engelsöy, T.G.~Mertens and H.~Verlinde, \emph{{An investigation of AdS$_{2}$ backreaction and holography}}, \href{https://doi.org/10.1007/JHEP07(2016)139}{\emph{JHEP} {\bfseries 07} (2016) 139} [\href{https://arxiv.org/abs/1606.03438}{{\ttfamily 1606.03438}}].

\bibitem{Jensen:2016pah}
K.~Jensen, \emph{{Chaos in AdS$_2$ Holography}}, \href{https://doi.org/10.1103/PhysRevLett.117.111601}{\emph{Phys. Rev. Lett.} {\bfseries 117} (2016) 111601} [\href{https://arxiv.org/abs/1605.06098}{{\ttfamily 1605.06098}}].

\bibitem{Kitaev:2017awl}
A.~Kitaev and S.J.~Suh, \emph{{The soft mode in the Sachdev-Ye-Kitaev model and its gravity dual}}, \href{https://doi.org/10.1007/JHEP05(2018)183}{\emph{JHEP} {\bfseries 05} (2018) 183} [\href{https://arxiv.org/abs/1711.08467}{{\ttfamily 1711.08467}}].

\bibitem{mehta2004random}
M.~Mehta, \emph{Random Matrices}, no.~142 in Pure and applied mathematics, Elsevier/Academic Press, ISBN:9780120884094 (2004).

\bibitem{Wigner:791e55682183}
E.P.~Wigner, \emph{Characteristic vectors of bordered matrices with infinite dimensions}, {\emph{Annals of Mathematics} {\bfseries 62} (1955) 548}.

\bibitem{You_2017}
Y.-Z.~You, A.W.W.~Ludwig and C.~Xu, \emph{Sachdev-ye-kitaev model and thermalization on the boundary of many-body localized fermionic symmetry-protected topological states}, \href{https://doi.org/10.1103/physrevb.95.115150}{\emph{Physical Review B} {\bfseries 95} (2017) }.

\bibitem{Garcia-Garcia:2016mno}
A.M.~Garc\'\i{}a-Garc\'\i{}a and J.J.M.~Verbaarschot, \emph{{Spectral and thermodynamic properties of the Sachdev-Ye-Kitaev model}}, \href{https://doi.org/10.1103/PhysRevD.94.126010}{\emph{Phys. Rev. D} {\bfseries 94} (2016) 126010} [\href{https://arxiv.org/abs/1610.03816}{{\ttfamily 1610.03816}}].

\bibitem{Dyer:2016pou}
E.~Dyer and G.~Gur-Ari, \emph{{2D CFT Partition Functions at Late Times}}, \href{https://doi.org/10.1007/JHEP08(2017)075}{\emph{JHEP} {\bfseries 08} (2017) 075} [\href{https://arxiv.org/abs/1611.04592}{{\ttfamily 1611.04592}}].

\bibitem{Cotler:2016fpe}
J.S.~Cotler, G.~Gur-Ari, M.~Hanada, J.~Polchinski, P.~Saad, S.H.~Shenker et~al., \emph{{Black Holes and Random Matrices}}, \href{https://doi.org/10.1007/JHEP05(2017)118}{\emph{JHEP} {\bfseries 05} (2017) 118} [\href{https://arxiv.org/abs/1611.04650}{{\ttfamily 1611.04650}}].

\bibitem{Gurau_2016}
R.G.~and, \emph{Invitation to random tensors}, \href{https://doi.org/10.3842/sigma.2016.094}{\emph{Symmetry, Integrability and Geometry: Methods and Applications} (2016) }.

\bibitem{Bonzom:2012hw}
V.~Bonzom, R.~Gurau and V.~Rivasseau, \emph{{Random tensor models in the large N limit: Uncoloring the colored tensor models}}, \href{https://doi.org/10.1103/PhysRevD.85.084037}{\emph{Phys. Rev. D} {\bfseries 85} (2012) 084037} [\href{https://arxiv.org/abs/1202.3637}{{\ttfamily 1202.3637}}].

\bibitem{Gurau:2011kk}
R.~Gurau, \emph{{Universality for Random Tensors}}, \href{https://doi.org/10.1214/13-AIHP567}{\emph{Ann. Inst. H. Poincare Probab. Statist.} {\bfseries 50} (2014) 1474} [\href{https://arxiv.org/abs/1111.0519}{{\ttfamily 1111.0519}}].

\bibitem{Bonzom_2013}
V.~Bonzom, R.~Gurau and M.~Smerlak, \emph{Universality in p-spin glasses with correlated disorder}, \href{https://doi.org/10.1088/1742-5468/2013/02/l02003}{\emph{Journal of Statistical Mechanics: Theory and Experiment} {\bfseries 2013} (2013) L02003}.

\bibitem{Krajewski:2018lom}
T.~Krajewski, M.~Laudonio, R.~Pascalie and A.~Tanasa, \emph{{Non-Gaussian disorder average in the Sachdev-Ye-Kitaev model}}, \href{https://doi.org/10.1103/PhysRevD.99.126014}{\emph{Phys. Rev. D} {\bfseries 99} (2019) 126014} [\href{https://arxiv.org/abs/1812.03008}{{\ttfamily 1812.03008}}].

\bibitem{Veneziano:1989hd}
G.~Veneziano, \emph{{Wormholes, Nonlocal Actions and a New Mechanism for Suppressing the Cosmological Constant}}, \href{https://doi.org/10.1142/S0217732389000824}{\emph{Mod. Phys. Lett. A} {\bfseries 4} (1989) 695}.

\bibitem{Coleman:1988tj}
S.R.~Coleman, \emph{{Why There Is Nothing Rather Than Something: A Theory of the Cosmological Constant}}, \href{https://doi.org/10.1016/0550-3213(88)90097-1}{\emph{Nucl. Phys. B} {\bfseries 310} (1988) 643}.

\bibitem{Polchinski:1988ua}
J.~Polchinski, \emph{{The Phase of the Sum Over Spheres}}, \href{https://doi.org/10.1016/0370-2693(89)90387-0}{\emph{Phys. Lett. B} {\bfseries 219} (1989) 251}.

\bibitem{Gross:2019uxi}
D.J.~Gross, J.~Kruthoff, A.~Rolph and E.~Shaghoulian, \emph{{Hamiltonian deformations in quantum mechanics, $T\bar T$, and the SYK model}}, \href{https://doi.org/10.1103/PhysRevD.102.046019}{\emph{Phys. Rev. D} {\bfseries 102} (2020) 046019} [\href{https://arxiv.org/abs/1912.06132}{{\ttfamily 1912.06132}}].

\bibitem{wigner_1951}
E.P.~Wigner, \emph{On the statistical distribution of the widths and spacings of nuclear resonance levels}, \href{https://doi.org/10.1017/S0305004100027237}{\emph{Mathematical Proceedings of the Cambridge Philosophical Society} {\bfseries 47} (1951) 790–798}.

\bibitem{Goel:2023svz}
A.~Goel, V.~Narovlansky and H.~Verlinde, \emph{{Semiclassical geometry in double-scaled SYK}},  \href{https://arxiv.org/abs/2301.05732}{{\ttfamily 2301.05732}}.

\bibitem{Das:2020kmt}
S.R.~Das, A.~Ghosh, A.~Jevicki and K.~Suzuki, \emph{{Near Conformal Perturbation Theory in SYK Type Models}}, \href{https://doi.org/10.1007/JHEP12(2020)171}{\emph{JHEP} {\bfseries 12} (2020) 171} [\href{https://arxiv.org/abs/2006.13149}{{\ttfamily 2006.13149}}].

\bibitem{Jevicki:2016bwu}
A.~Jevicki, K.~Suzuki and J.~Yoon, \emph{{Bi-Local Holography in the SYK Model}}, \href{https://doi.org/10.1007/JHEP07(2016)007}{\emph{JHEP} {\bfseries 07} (2016) 007} [\href{https://arxiv.org/abs/1603.06246}{{\ttfamily 1603.06246}}].

\bibitem{Jevicki:2016ito}
A.~Jevicki and K.~Suzuki, \emph{{Bi-Local Holography in the SYK Model: Perturbations}}, \href{https://doi.org/10.1007/JHEP11(2016)046}{\emph{JHEP} {\bfseries 11} (2016) 046} [\href{https://arxiv.org/abs/1608.07567}{{\ttfamily 1608.07567}}].

\end{thebibliography}\endgroup

\end{document}